\documentclass[twocolumn,nofootinbib,preprintnumbers,floatfix,prd,altaffilletter]{revtex4-1}

\usepackage[bookmarks=false,hyperfootnotes=false]{hyperref}
\usepackage{graphicx}
\usepackage{amsmath}
\usepackage{xspace}
\usepackage{xcolor}
\usepackage{float}
\usepackage{comment}
\usepackage{subcaption}
\usepackage{textcomp}
\usepackage{placeins}
\usepackage[normalem]{ulem}
\usepackage{lipsum}
\usepackage{epsf,amssymb}
\usepackage{booktabs,tabularx}
\usepackage{rotating}
\usepackage{microtype}
\usepackage[titletoc,title]{appendix}
\usepackage{amsmath,amsfonts,amsthm,bm} 
\usepackage[tikz]{bclogo}
\usepackage[compat=1.1.0]{tikz-feynman}
\usepackage{xspace}
\usepackage{wasysym}

\newcommand{\Eq}[1]{Eq.~\eqref{eq:#1}}

\newcommand{\Eqstwo}[2]{Eqs.~\eqref{eq:#1} and \eqref{eq:#2}}

\newcommand{\Sec}[1]{Sec.~\ref{sec:#1}}

\newcommand{\Fig}[1]{Fig.~\ref{fig:#1}}

\newcommand{\citeRef}[1]{\mbox{Ref.~\cite{#1}}}

\newcommand{\citeRefs}[1]{\mbox{Refs.~\cite{#1}}}

\newcommand{\sphid}[1]{}

\providecommand{\href}[2]{#2}

\newcommand\as{\alpha_{\mathrm{S}}}

\def\to{\rightarrow}

\def\nn{\nonumber}

\newcommand\Matrix{{\sc Matrix}\xspace}

\newcommand{\pt}{\ensuremath{q_\perp}}

\newcommand{\rd}{\mathrm{d}} 
\newcommand{\re}{e}

\newcommand{\df}{\mathrm{d}}

\newcommand{\dsigMC}{\df\sigma^\textsc{mc}}

\newcommand{\obs}{X}
\newcommand{\Tau}{\mathit r}
\newcommand{\cut}{\mathrm{cut}}
\newcommand{\cP}{\mathcal{P}}
\newcommand{\nons}{\mathrm{nons}}
\newcommand{\eps}{\delta}
\newcommand{\tildeR}{R}
\newcommand{\dZ}{\rd\mathcal{Z}[\{{\tildeR}', k_i\}]}

\newcommand{\kto}{k_{\perp,1}}

\newcommand{\tildeL}{{L}}
\newcommand{\tildecalL}{{\cal L}}
\newcommand{\tildeC}{C}
\newcommand{\tildeH}{H}

\newcommand{\Ecm}{E_{\rm cm}}

\newcommand{\geneva}{\textsc{Geneva}\xspace}
\newcommand{\genevapt}{\textsc{Geneva}$_{q_\perp}$\xspace}
\newcommand{\genevatau}{\textsc{Geneva}$_{\mathcal T_0}$\xspace}

\newcommand{\pythia}{\textsc{Pythia}\xspace}
\newcommand{\pythiaEight}{\textsc{Pythia8}\xspace}

\newcommand{\radish}{\textsc{RadISH}\xspace}
\newcommand{\zerojet}{0-jet\xspace}
\newcommand{\onejet}{1-jet\xspace}

\allowdisplaybreaks[1]

\newcommand{\rescaleoneplot}{0.8\columnwidth}
\newcommand{\rescaletwoplots}{0.41\textwidth}
\newcommand{\hspacebetweentwoplots}{1em}
\newcommand{\rescalethreeplots}{0.32\textwidth}
\newcommand{\hspacebetweenthreeplots}{\fill}

\newcommand{\spacebeforefigurecaption}{-1ex}

\begin{document}


\title{Matching NNLO to parton shower using N$^3$LL colour-singlet\\ transverse momentum resummation in G{\footnotesize ENEVA}}

\preprint{ZU-TH 8/21}
\author{Simone~Alioli}
\email{simone.alioli@unimib.it}
\author{Alessandro~Broggio}
\email{alessandro.broggio@unimib.it}
\author{Alessandro~Gavardi}
\email{a.gavardi@campus.unimib.it}
\author{Stefan~Kallweit}
\email{stefan.kallweit@unimib.it}
\author{Matthew~A.~Lim}
\email{matthew.lim@unimib.it}
\author{Riccardo~Nagar}
\email{riccardo.nagar@unimib.it}
\author{Davide~Napoletano}
\email{davide.napoletano@unimib.it}
\affiliation{Universit\`{a} degli Studi di Milano-Bicocca \& INFN, Piazza della Scienza 3, Milano 20126, Italy}
\author{Christian~W.~Bauer}
\email{cwbauer@lbl.gov}
\affiliation{Physics Division, Lawrence Berkeley National Laboratory, Berkeley, CA 94720, USA}
\author{Luca~Rottoli}
\email{luca.rottoli@physik.uzh.ch}
\affiliation{Physik Institut, Universit\"at Z\"urich, CH-8057 Z\"urich, Switzerland}
\date{\today}

\begin{abstract}
   We extend the \geneva Monte Carlo framework using the transverse momentum of a colour-singlet system as the resolution variable.
   This allows us to use next-to-next-to-next-to leading logarithm (N$^3$LL) resummation via the \radish formalism to obtain precise predictions for any colour-singlet production process at the fully exclusive level.
   Thanks to the implementation of two different resolution variables within the \geneva framework, we are able to assess the impact of such a choice on  differential observables for the first time.
   As a first application we present predictions for Drell-Yan lepton pair production at next-to-next-to-leading order (NNLO) in QCD interfaced to a parton shower simulation that includes additional all-order radiative corrections.
   We provide fully showered and hadronised events using \pythiaEight, while retaining the NNLO QCD accuracy for observables which are inclusive over the additional radiation.     
   We also show that it is possible to obtain a numerically good agreement between showered \geneva predictions and the N$^3$LL resummation for the transverse momentum spectrum by choosing a more local recoil scheme. 
      We compare our final predictions to LHC data at 13 TeV, finding good agreement across several distributions.

\end{abstract}
\keywords{QCD,NNLO,Parton Shower, Resummation}

\maketitle
\flushbottom

\section{Introduction}
\label{sec:intro}

The rich and vast physics program at the Large Hadron Collider (LHC) has delivered an outstanding amount of data so far.
Thanks to these impressive performances, particle physics has entered an era where high precision is ubiquitous. With the forthcoming start of the Run3 and the future upgrade to the High-Luminosity LHC, the amount of data that will be collected  will increase significantly. This groundbreaking machine will thus be able to detect extremely rare phenomena and further improve the precision of measurements.

Theoretical predictions of the Standard Model (SM) processes are one of the most fundamental ingredients for the interpretation of collider data. The vast majority of experimental analyses rely on  theoretical predictions in the form of perturbative calculations at parton level or in combination with parton showers (PS) in fully exclusive Monte Carlo (MC) event generators.
The advantage of MC event generators is the ability to produce hadron-level events that can be directly interfaced to detector simulations.

In order to completely exploit the precision of present and future experimental data, it becomes therefore mandatory to refine MC event generators including the state-of-the-art corrections available.

For this reason, in recent years there has been an increasing interest in obtaining theoretical predictions where fixed-order predictions  at next-to-next-to-leading order (NNLO) accuracy are consistently matched with parton shower simulations.
There are currently four available methods which can reach NNLO+PS accuracy~\cite{Hamilton:2013fea,Alioli:2013hqa,Alioli:2015toa,Hoeche:2014aia,Hoche:2014dla,Monni:2019whf,Monni:2020nks}, which have been initially applied to $2 \rightarrow 1$ processes such as Higgs~\cite{Hamilton:2013fea,Hoche:2014dla} and Drell-Yan~\cite{Karlberg:2014qua,Hoeche:2014aia,Alioli:2015toa} production.
The number of applications at hadron colliders has been rapidly growing in the last few years, and the methods have been extended to more complex processes such as Higgsstrahlung~\cite{Astill:2016hpa,Astill:2018ivh,Alioli:2019qzz}, hadronic Higgs boson decays~\cite{Bizon:2019tfo,Alioli:2020fzf}, $W^+W^-$~\cite{Re:2018vac}, diphoton~\cite{Alioli:2020qrd}, $Z \gamma$~\cite{Lombardi:2020wju}, and $t \bar t$ production~\cite{Mazzitelli:2020jio}.

In this paper we focus on the \geneva framework, which has been developed in \citeRefs{Alioli:2012fc,Alioli:2013hqa,Alioli:2015toa,Alioli:2016wqt}.
This method attains NNLO+PS accuracy by combining fixed-order predictions at NNLO accuracy with the higher-order resummation of an $N$-jet resolution variable and matching the ensuing predictions with the parton shower.
The $N$-jet resolution variable partitions the phase space into regions with a different number of resolved emissions in the final state, such that infrared (IR) divergent final states with $M$ partons are translated into finite final states with $N$ partonic jets (where $M \geq  N$).
This ensures that the IR divergences cancel on an event-by-event basis yielding physical and IR-finite events.

Although the \geneva method is completely general to the point that several different resolution variables can be used, so far the only practical implementations employed $N$-jettiness  $\mathcal T_N$~\cite{Stewart:2010tn} to achieve the separation between the $N$ and $N+1$ jet events.
Nonetheless, any other resolution variable which can be resummed at high enough accuracy can be used.
In this paper, we extend the \geneva method using the transverse momentum $\pt$ of a colour-singlet system as a \zerojet separation variable.
The choice of this variable is principally dictated by the availability of higher-order resummation, up to the next-to-next-to-next-to-leading logarithm (N$^3$LL), and by the extreme precision at which it is measured by the LHC experiments, for different processes.

However, the peculiar vectorial nature of this observable deserves further comment, especially in view of its usage as \zerojet resolution.
Indeed, at variance with $N$-jettiness, it does not completely single out the soft and collinear limits when $\pt \rightarrow 0$.
The reason is that there are two competing mechanisms which can drive $\pt$ to
small values: the first is the usual ensemble of soft and/or collinear
partonic emissions with $k_{\perp,i} \simeq 0$ recoiling against the produced
colour singlet; the second proceeds instead through a combination of two or more relatively hard emissions balancing each other, such that the resulting vectorial sum of their $\vec k_{t,i}$ is zero.
It happens that the latter mechanism is the one that dominates at small $\pt$ and is responsible for the behaviour of the differential cross-section in the small $\pt$ limit: indeed, in the low $\pt$ region the spectrum vanishes as $d\sigma /d\pt \sim \pt$, instead of vanishing exponentially as the Sudakov suppression of the first mechanism would suggest~\cite{Parisi:1979se}.

The presence of these two competing mechanisms seems to prevent the usage of  $\pt$ as a \zerojet resolution observable.
However, both effects which lead to a vanishing $\pt$ are properly included in the \zerojet cross section by performing the resummation of $\pt$ in the Fourier-conjugate impact-parameter space~\cite{Collins:1984kg}.
The problem of resummation in direct space is more delicate, and only recently was it shown that it is possible to directly resum the transverse momentum spectrum in direct space without spoiling the scaling at low $\pt$~\cite{Monni:2016ktx,Bizon:2017rah}.
This problem was also addressed in \citeRef{Ebert:2016gcn} within a Soft-Collinear Effective Theory (SCET) approach, where the renormalisation- group evolution is solved directly in momentum space.
The remaining nonsingular contribution stemming from two or more relatively hard emissions resulting in a system with small $\pt$ are strongly suppressed, and can be neglected.
As a result it is possible to use $\pt$ as a proper \zerojet resolution variable.
In particular, in this work we consider the \radish approach of \citeRefs{Monni:2016ktx,Bizon:2017rah}, which allows us to resum the transverse momentum spectrum at N$^3$LL.

Here we focus on neutral Drell-Yan (DY) production at the LHC ($pp \rightarrow Z/\gamma^* \rightarrow \ell^+ \ell^-$) as a case study, but this approach can immediately be applied also to the charged current case and to other colour-singlet processes.
Differential distributions of electroweak gauge bosons play a paramount role in the precision programme at the LHC.
These observables are measured at the level of their leptonic decays and are typically characterised by particularly small experimental uncertainties, which can reach the few permille level in neutral DY production~\cite{Chatrchyan:2011wt,Aad:2012wfa,Aad:2014xaa,Aad:2015auj,Aaij:2015gna,Khachatryan:2015oaa,Khachatryan:2015paa,Aaij:2015zlq,Aad:2016izn,Khachatryan:2016nbe,Aaij:2016mgv,Sirunyan:2017igm,Aaboud:2017svj,Aaboud:2017ffb,Aad:2019wmn,Sirunyan:2019bzr}.
This further motivates the inclusion of higher-order calculations in MC event generators for this process.

On the theory side, fixed-order predictions for neutral DY production have been known at NNLO accuracy for quite some time at the inclusive~\cite{Hamberg:1990np,vanNeerven:1991gh,Anastasiou:2003yy,Melnikov:2006di,Melnikov:2006kv,Catani:2010en,Catani:2009sm,Gavin:2010az,Anastasiou:2003ds} and at the differential level~\cite{Ridder:2015dxa,Ridder:2016nkl,Gehrmann-DeRidder:2016jns,Gauld:2017tww,Boughezal:2015ded,Boughezal:2016isb}.
 Due to the outstanding precision of the experimental data the theoretical calculations are currently being pushed at next-to-next-to-next-to-leading order (N$^3$LO), and the inclusive cross section for lepton-pair production through a virtual photon has been recently calculated at this accuracy in \citeRef{Duhr:2020seh}.
 Electroweak (EW) corrections for this process are known~\cite{Baur:1997wa,Baur:2001ze,Kuhn:2005az,Zykunov:2005tc,Arbuzov:2007db,CarloniCalame:2007cd,Dittmaier:2009cr,Denner:2011vu,Buonocore:2019puv}.
  The computation of mixed QCD--EW corrections is an active area of research~\cite{Dittmaier:2014qza,Dittmaier:2015rxo,deFlorian:2018wcj,Delto:2019ewv,Bonciani:2019nuy,Cieri:2020ikq}, and they were recently computed for the production of an on-shell $Z$ boson and its decay to massless charged leptons has become available~\cite{Buccioni:2020cfi}.

 In this work, we improve the fully differential NNLO calculation with the N$^3$LL resummation for the transverse momentum and we subsequently shower and hadronise the events with \pythiaEight~\cite{Sjostrand:2014zea}, while maintaining NNLO accuracy for the underlying process and a numerically good agreement with the resummed $\pt$ spectrum.
We compare our predictions with recent data collected at the LHC by the ATLAS~\cite{Aad:2019wmn} and CMS~\cite{Sirunyan:2019bzr} collaborations at a centre-of-mass energy of 13~TeV.

The manuscript is organised as follows.
In \Sec{theory} we introduce the theoretical framework; we first review the \geneva method by discussing its extension to a different \zerojet resolution variable and recap the resummation of the transverse momentum within the \radish formalism.
In \Sec{implementation} we discuss the details of the implementation and present the validation of our results by performing comparisons against fixed-order predictions at NNLO.
We also study the differences with respect to the original \geneva implementation of \citeRefs{Alioli:2015toa,Alioli:2016wqt}, which uses the beam thrust as the \zerojet resolution variable.
We then compare our predictions with the experimental data in \Sec{result}
and finally draw our conclusions in \Sec{conclusion}.
  
\section{Theoretical framework}\label{sec:theory}

In this section we briefly review the theoretical framework used in this work.
We start by providing a short description of the \geneva method, focussing especially on the separation between 0- and \onejet events. This is particularly relevant in our case due to the change in the \zerojet resolution variable.
We then discuss the resummation of the transverse momentum spectrum in direct space using the \radish formalism.

\subsection{The G{\scriptsize ENEVA} method}\label{sec:geneva}

The \geneva method is based on the definition of physical and IR-finite events at a given perturbative accuracy, with the condition that IR singularities cancel on an event-by-event basis. This is  achieved  by mapping IR-divergent final states with $M$ partons into IR-finite final states with $N$ jets, with $M \geq N$.
Events are classified according to the value of $N$-jet resolution variables $\Tau_N$ which partition the phase space into different regions according to the number of resolved emissions.
In particular, the \geneva Monte Carlo cross section $\dsigMC_N$ receives
contributions from both $N$-parton events and $M$-parton events where the additional emission(s) are below the resolution cut $\Tau_N^\cut$ used to separate resolved and unresolved emissions.
The unphysical dependence on the boundaries of this partitioning procedure is removed by requiring that the resolution parameters are resummed at high enough accuracy.
Considering events with up to two resolved emissions, necessary to achieve NNLO accuracy for colour-singlet production, one has
\begin{align}
\label{eq:NNLOevents}
\text{$\Phi_0$ events: }
& \qquad \frac{\dsigMC_0}{\df\Phi_0}(\Tau_0^\cut)
\,,\nn\\
\text{$\Phi_1$ events: }
& \qquad
\frac{\dsigMC_{1}}{\df\Phi_{1}}(\Tau_0 > \Tau_0^\cut; \Tau_{1}^\cut)
\,,
\\
\text{$\Phi_2$ events: }
& \qquad
\frac{\dsigMC_{\ge 2}}{\df\Phi_{2}}(\Tau_0 > \Tau_0^\cut, \Tau_{1} > \Tau_{1}^\cut)
\nn \,,
\end{align}
where by using the notation $\Tau_N > \Tau_N^\cut$ we indicate that the events
are differential in the $N$-jet resolution variable, whilst with the notation
$\Tau_N^\cut$ we indicate that the MC cross section contains events in which the
resolution variable has been integrated up to the resolution cut.
As mentioned in the introduction, the definition of the partonic jet bins is based on the definition of a suitable phase space mapping $\Phi_N(\Phi_M)$, where $N$ and $M $ denote the number of jets and partons respectively. This ensures the finiteness of the $\dsigMC_N$ cross section; however, they must not be mistaken with the jet bins commonly used in experimental analyses, which instead define jets according to a particular jet algorithm.\footnote{In principle nothing prevents the usage of a resolution variable based on a standard jet-algorithm to define the \geneva jet bins, e.g.~$p_{T,\rm jet}$, if not for the lack of the corresponding higher-order logarithmic resummation.}

Defining the events as in \Eq{NNLOevents} the cross section for a generic observable $X$ is
\begin{align}
\sigma(\obs)
&= \int\!\df\Phi_0\, \frac{\dsigMC_0}{\df\Phi_0}(\Tau_0^\cut)\, M_\obs(\Phi_0) \nn \\ & \quad
+ \int\!\df\Phi_{1}\, \frac{\dsigMC_{1}}{\df\Phi_{1}}(\Tau_0 > \Tau_0^\cut; \Tau_{1}^\cut)\, M_\obs(\Phi_{1})
\\ & \quad
+ \int\!\df\Phi_{2}\, \frac{\dsigMC_{\ge 2}}{\df\Phi_{2}}(\Tau_0 > \Tau_0^\cut, \Tau_{1} > \Tau_{1}^\cut)\, M_\obs(\Phi_{2}) \nn
\,,
\end{align}
where by $M_\obs(\Phi_N)$ we indicate the measurement function used to compute the observable $\obs$ for the $N$-jet final state $\Phi_N$.
We stress that the expression $\sigma(\obs)$ is not exactly equivalent to the result of a fixed-order calculation at NNLO accuracy as for unresolved emissions the observable is calculated on the projected phase space $\Phi_N (\Phi_M)$ rather than on $\Phi_M$.
However, the difference is of nonsingular nature and vanishes in the limit $\Tau_N^\cut \rightarrow 0$.
The value of the resolution cut $\Tau_N^\cut$ should therefore be chosen as small as possible.
In this limit, the result contains large logarithms of the resolution variable $\Tau_N$ (and of $\Tau_N^\cut$) which one should resum to all orders in perturbation theory to yield meaningful results.

Let us start by discussing the separation between the \zerojet and the \onejet regimes and the associated resummation of the \zerojet resolution variable.
The expression for the MC cross sections in the exclusive \zerojet bin and in the inclusive \onejet bin are
\begin{align}
\frac{\dsigMC_0}{\df\Phi_0}(\Tau_0^\cut)
&= \frac{\df\sigma^{\rm res}}{\df\Phi_0}(\Tau_0^\cut)
+ \frac{\df\sigma_0^{\rm nons}}{\df\Phi_0}(\Tau_0^\cut)
\,,
\label{eq:0master}
\end{align}
and
\begin{align}
\frac{\dsigMC_{\geq 1}}{\df\Phi_{1}}(\Tau_0 > \Tau_0^\cut)
&= \frac{\df\sigma^{\rm res}}{\df\Phi_{0}\df \Tau_0} \, \cP(\Phi_1) \theta\left(\Tau_0 > \Tau_0^\cut\right)
\label{eq:1incmaster} \nn\\ &\qquad + \frac{\df\sigma_{\ge 1}^{\rm nons}}{\df\Phi_1}(\Tau_0 > \Tau_0^\cut)
\,,
\end{align}
respectively. The label `$\rm res$' stands for `resummed' and the nonsingular contributions labelled `$\rm nons$' must only contain terms which are nonsingular in the $\Tau_0 \rightarrow 0$ limit.
When working at NNLO accuracy, to ensure that the second contribution on the right hand side of \Eqstwo{0master}{1incmaster} is genuinely nonsingular, the resummed contribution must include all the terms singular in the resolution variable $\Tau_0$ at order $\alpha_s^2$, i.e.~it should be evaluated at least at NNLL$^\prime$ accuracy.
In the context of hadron collider processes, the \geneva implementations have so far only employed  the beam thrust $\mathcal T_0$  as \zerojet resolution variable, whose resummation was performed at NNLL$^\prime$ in SCET.
In this work we extend the \geneva framework to use the transverse momentum of the colour-singlet $\pt$ as \zerojet resolution parameter, which we resum at N$^3$LL accuracy, see paragraph~\ref{sec:resumm-radish}.
It is worth emphasising again that when one chooses $\pt$ as the  \zerojet resolution, one is not only singling out configurations without any hard emission as there exist configurations with two or more partons with transverse momenta balancing each other such that the vectorial sum of their $k_\perp$ is small.
These contributions are included by the resummation in $\frac{\dsigMC_0}{\df\Phi_0}(\pt^{\rm cut})$, but they are only described in the soft and/or collinear approximation, missing power suppressed nonsingular corrections which are progressively important at increasing values of $k_\perp$.
However, as long as one keeps $\pt^{\rm cut}$ sufficiently small, the chance of obtaining a small vectorial sum from larger and larger individual transverse momenta is heavily suppressed, and therefore the contribution to this region stemming from hard jets can be safely neglected.
We will discuss the value of $\pt^{\rm cut}$ necessary to make the power-suppressed terms sufficiently small that they may be neglected in Sect.~\ref{sec:power-suppressed-terms}.

In order to make the $\Tau_0$ spectrum fully differential in the $\Phi_1$ phase space, \Eq{1incmaster} contains a splitting function $\cP(\Phi_1)$ fulfilling the normalisation condition
\begin{align}
\label{eq:Pnorm}
\int \! \frac{\df\Phi_1}{\df \Phi_{0} \df \Tau_0} \, \cP(\Phi_1) = 1
\,.\end{align}
This function is used to extend the differential dependence of $\df\sigma^{\rm res}$ including  the full radiation phase space, written as a function of $\Tau_0$  and two additional variables, which we choose to be the energy fraction $z$ and the azimuthal angle $\phi$  of the emission.

The nonsingular contributions entering in \Eqstwo{0master}{1incmaster} are
\begin{align}
\frac{\df\sigma_0^\nons}{\df\Phi_{0}}&(\Tau_0^\cut)
\label{eq:0nons}
= \frac{\df\sigma_0^{{\rm NNLO_0}}}{\df\Phi_{0}}(\Tau_0^\cut)
- \biggl[\frac{\df\sigma^{\rm res}}{\df\Phi_{0}}(\Tau_0^\cut) \biggr]_{\rm NNLO_0}
\end{align}
and
\begin{align}
\frac{\df\sigma_{\ge 1}^\nons}{\df\Phi_{1}}&(\Tau_0 > \Tau_0^\cut)
\label{eq:1nons} =  \frac{\df\sigma_{\ge 1}^{{\rm NLO_1}}}{\df\Phi_{1}}(\Tau_0 > \Tau_0^\cut)  \\ &\quad - \biggl[\frac{\df\sigma^{\rm res}}{\df\Phi_0 \df \Tau_0}\cP(\Phi_1) \biggr]_{\rm NLO_1} \!\!\!\,\theta\left(\Tau_0 > \Tau_0^\cut\right)
\,,\nn
\end{align}
where NNLO$_0$ and NLO$_1$ indicate the accuracy at which one should compute the fixed-order contributions for each jet bin. The terms in square brackets are the expansion of the resummed result at $\mathcal O(\as^2)$ of the cumulant and of the spectrum.

By writing explicitly the expressions for the fixed-order cross sections one has
\begin{align}
\frac{\dsigMC_0}{\df\Phi_0}&(\Tau_0^\cut) =
\frac{\df\sigma^{\rm res}}{\df\Phi_0}(\Tau_0^\cut)\, - \biggl[\frac{\df\sigma^{\rm res}}{\df\Phi_{0}}(\Tau_0^\cut) \biggr]_{\rm NNLO_0} \, \nn\\
&+(B_0+V_0+W_0)(\Phi_0)  \nn \\
&+  \int \frac{\mathrm{d} \Phi_1}{\mathrm{d} \Phi_0} (B_1 + V_1)(\Phi_1)\,\theta\big( \Tau_0(\Phi_1)< \Tau_0^{\mathrm{cut}}\big)\, \nn \\
&+  \int \frac{\mathrm{d} \Phi_2}{\mathrm{d} \Phi_0} \,B_2 (\Phi_2)\, \theta\big( \Tau_0(\Phi_2)< \Tau_0^{\mathrm{cut}}\big)\, ,
\label{eq:0full}
\end{align}
for the 0-jet bin, and
\begin{align}
\label{eq:sigma>=1}
\frac{\dsigMC_{\geq 1}}{\df\Phi_{1}}&(\Tau_0 > \Tau_0^\cut) = \nn\\
&\Bigg\{\frac{\df\sigma^{\rm res}}{\df\Phi_{0}\df \Tau_0} -  \biggl[\frac{\df\sigma^{\rm res}}{\df\Phi_0 \df \Tau_0} \biggr]_{\rm NLO_1}\Bigg\}\, \cP(\Phi_1)\, \theta\left(\Tau_0 > \Tau_0^\cut\right) \nn\\&
  +(B_1+V_1)(\Phi_1) \theta (\Tau_0(\Phi_1)> \Tau_0^{\mathrm{cut}}) \nn\\& + \int  \frac{\df\Phi_{2}}{\df\Phi^{\Tau_0}_1}\, B_2(\Phi_2)\,  \theta\big( \Tau_0(\Phi_2) > \Tau_0^{\mathrm{cut}}\big)
\end{align}
for the inclusive 1-jet bin.

In the above equations $B_M$ are the $M$-parton tree-level contributions, $V_M$ are the $M$-parton one-loop contributions, and finally $W_0$ is the two-loop contribution.
In \Eq{sigma>=1} we also introduced the notation
\begin{align}
\label{eq:dPhiRatio}
\frac{\df \Phi_{2}}{\df \Phi^{\Tau_0}_1} \equiv \df \Phi_{2} \, \delta[ \Phi_1 - \Phi^{\Tau_0}_1(\Phi_2)] \Theta^{\Tau_0}(\Phi_1)
\end{align}
to indicate that the integration is performed  over a region of $\Phi_2$ while keeping the values of $\Tau_0$ and of the $\Phi_1$ observables that we use to parametrise the \onejet phase space fixed. 
This is required since the expression for $\dsigMC_{\geq 1}$ is differential in the \zerojet resolution parameter $\Tau_0$, meaning that we should also parametrise the 2-parton contribution $B_2$ in the same terms.
This means that the mapping used in the projection $\df\Phi_{2}/\df\Phi^{\Tau_0}_1$ should preserve $\Tau_0$, i.e.
\begin{equation} \label{eq:Tau0map}
\Tau_0(\Phi_1^{\Tau_0}(\Phi_2)) = \Tau_0(\Phi_2)\,,
\end{equation}
to guarantee the pointwise cancellation of the singular $\Tau_0$ contributions in \Eq{sigma>=1}.

The $\Theta^{\Tau_0}(\Phi_1)$ in \Eq{dPhiRatio} limits the integration to the singular points for which it is possible to construct such a projection.
The non-projectable region of the $\Phi_2$ phase space, which only
includes nonsingular events, must therefore be included in the 2-jet event
cross section, as shown later.
We finally notice that \Eq{sigma>=1} should be supplemented by including in $\dsigMC_{1}$ \onejet events with $\Tau_0 <  \Tau_0^{\rm cut}$ which cannot be mapped into \zerojet events.
In the case of the $1 \rightarrow 0$ case we use the FKS mapping~\cite{Frixione:1995ms}, which implies that the only non-projectable events are those which would result in an invalid flavour configuration that is not present at the LO level (e.g.~$qg \rightarrow \ell^+ \ell^- q$ projected to $gg \rightarrow \ell^+ \ell^-$).
By denoting the non-projectable region with the symbol $\overline{\Theta}_{\rm map}^{\rm FKS}$, we supplement the formul\ae\ above with
\begin{align}\label{eq:tau0smaller}
\frac{\dsigMC_{ 1}}{\df\Phi_{1}}(\Tau_0 \leq \Tau_0^\cut) &=  (B_1+V_1)\, (\Phi_1)\,\theta(\Tau_0<\Tau^{\mathrm{cut}}_0)\,  \overline{\Theta}^{\mathrm{FKS}}_{\mathrm{map}}(\Phi_1)  \, .
\end{align}

Before we move on to discuss the N$^3$LL resummation of the transverse momentum, we shall briefly review the separation of the \onejet cross section into an exclusive \onejet cross section and an inclusive 2-jet cross section.
This separation can be performed in analogy to the 0-/1-jet separation discussed above, with $\Tau_1$ now being the relevant resolution variable:
\begin{align}
\frac{\dsigMC_{1}}{\df\Phi_{1}}  (\Tau_0 > & \Tau_0^\cut;  \Tau_{1}^\cut)
  =  \frac{\df\sigma_1^{\rm res}}{\df\Phi_{1}}(\Tau_0 > \Tau_0^\cut;  \Tau_{1}^\cut)
\nn  \\&\ + \frac{\df\sigma_1^{\mathrm{match}}}{\df \Phi_{1}}(\Tau_0 > \Tau_0^\cut; \Tau_{1}^\cut)
\,,  \label{eq:1master}
\\
\frac{\dsigMC_{\geq 2}}{\df\Phi_{2}}  (\Tau_0 > \Tau_0^\cut,& \Tau_{1}>\Tau_{1}^\cut)
 = \frac{\df\sigma^{\rm res}_{\geq 2}}{\df\Phi_2} \left(\Tau_0 > \Tau_0^\cut , \Tau_{1} > \Tau_{1}^\cut \right) \nn \\ &\
+ \frac{\df \sigma^{\mathrm{match}}_{\geq 2}}{\df \Phi_2}(\Tau_0 > \Tau_0^\cut, \Tau_1 > \Tau_1^\cut)
\,.
\label{eq:2master}
\end{align}
In these equations the resummation accuracy can be lowered with respect to the 0-/1-jet separation and is set to NLL.
The full expressions for the exclusive \onejet and the inclusive 2-jet cross section read~\cite{Alioli:2015toa,Alioli:2019qzz}
\begin{widetext}
\begin{align} \label{eq:1masterfull}
\frac{\dsigMC_{1}}{\df\Phi_{1}} (\Tau_0 > \Tau_0^\cut; \Tau_{1}^\cut)
&=\Bigg\{ \frac{\df\sigma^{\rm res}}{\df\Phi_0\df\Tau_0}\, \cP(\Phi_1)-\Bigg[\frac{\df\sigma^{\rm res}}{\df\Phi_0\df\Tau_0} \cP(\Phi_1)\Bigg]_{{\rm NLO}_1} + \big[B_1 + V_1^C\big](\Phi_1)  \Bigg\} \times  U_1(\Phi_1, \Tau_1^\cut)\, \theta(\Tau_0 > \Tau_0^\cut)
\nn \\ &
\quad +\bigg\{\int\ \biggl[\frac{\df\Phi_{2}}{\df\Phi^{\Tau_0}_1}\,B_{2}(\Phi_2)\, \theta\left(\Tau_0(\Phi_2) > \Tau_0^\cut\right)\,\theta(\Tau_{1} < \Tau_1^\cut) - \frac{\df\Phi_2}{\df \Phi_1^C}\, C_{2}(\Phi_{2})\, \theta(\Tau_0 > \Tau_0^\cut) \biggr]\bigg\}
\nn \\ & \quad - B_1(\Phi_1)\, U_1^{(1)}(\Phi_1, \Tau_1^\cut)\ \theta(\Tau_0 > \Tau_0^\cut)\, ,
\\
\label{eq:2masterful}
\frac{\dsigMC_{\geq 2}}{\df\Phi_{2}} (\Tau_0 > \Tau_0^\cut, \Tau_{1}>\Tau_{1}^\cut)  &= \left[
\Bigg\{ \frac{\df\sigma^{\rm res}}{\df\Phi_0\df\Tau_0} \, \cP(\Phi_1) -  \bigg[\frac{\df\sigma^{\rm res}}{\df\Phi_0\df\Tau_0}\, \cP(\Phi_1) \bigg]_{{\rm NLO}_1}\,  + (B_1 + V_1^C)(\Phi_1)\Bigg\} \,  U_1'(\Phi_1, \Tau_1) \right.
\nn \\ &\quad \left. \phantom{\Bigg\{ } \times
\theta(\Tau_0 > \Tau_0^\cut)  \right]_{{\Phi}_1 = \Phi_1^{\Tau_0}(\Phi_2)} \!\!\cP(\Phi_2) \, \theta(\Tau_1 > \Tau_1^\cut)
\nn \\ &\quad
+ \big\{ B_2(\Phi_2)\, \theta(\Tau_{1}>\Tau^{\mathrm{cut}}_{1})
 - B_1(\Phi_1^{\Tau_0})\,U_1^{(1)\prime}(\Phi_1^{\Tau_0}, \Tau_1)\,\cP(\Phi_2)\,\theta(\Tau_1 > \Tau_1^\cut)
\big\} \times \theta\left(\Tau_0(\Phi_2) > \Tau_0^\cut\right)
\,.\end{align}
\end{widetext}
In these formul\ae\, $U_1 (\Phi_1, \Tau_1^{\cut})$ is the Sudakov form factor that resums the dependence on $ \Tau_1^{\cut}$ at NLL, $  U_1'(\Phi_1, \Tau_1)$ denotes its derivative with respect to $ \Tau_1^{\cut}$, and $U_1^{(1)}$ and $U_1^{(1)\prime}$ are their $\mathcal O(\as)$ expansions.
The $C_2$ term is the singular approximant of the double-real matrix element $B_2$, which acts as a standard NLO subtraction reproducing the singular behaviour of $B_2$.
The term $V_1^C$ is defined from the relation
\begin{align}
	(B_1 + V_1) (\Phi_1) + \int \frac{\df \Phi_2}{\df \Phi_1^C} C_2 (\Phi_2) \equiv (B_1 + V_1^C) (\Phi_1)\,.
\end{align}
The normalised splitting function $\cP (\Phi_2)$ satisfies the unitarity condition (cfr.~the 0-/1-jet case \Eq{Pnorm})
\begin{align}\label{eq:P2norm}
	U_1 (\Phi_1, \Tau_1^\cut ) + \int \frac{\df \Phi_2}{\df \Phi_1^{\Tau_0}} \, U_1^\prime (\Phi_1, \Tau_1) \cP (\Phi_2) \, \theta (\Tau_1 > \Tau_1^\cut) = 1.
\end{align}
Analogously to the case of the normalised splitting function  $\cP (\Phi_1)$,  the function $\cP (\Phi_2)$ extends the differential dependence of $\df\sigma^{\rm res}$ including the radiation phase space for two emissions, where the second emission  should now be parametrised in terms of $\Tau_1$  and of the same two additional radiation variables.
Despite having used the same notation, we stress that the mapping used in \Eq{2masterful} and \Eq{P2norm} does not need to correspond to the one used to implement the subtraction in \Eq{1masterfull}. Indeed the latter does not need to be written as a function of $\Tau_1$.

The formul\ae\ above should be supplemented by including non-projectable 2-jet events with $\Tau_1 < \Tau_1^\cut$, namely
\begin{align}\label{eq:nonsing2jet}
\frac{\dsigMC_{\geq 2}}{\df\Phi_{2}} & (\Tau_0 > \Tau_0^\cut, \Tau_{1} \le \Tau_{1}^\cut)
\, = \, \nn \\& B_2(\Phi_2)\, \overline{\Theta}_{\mathrm{map}}^{\Tau_0}(\Phi_2)\,\theta(\Tau_1 < \Tau_1^\cut)\, \theta\left(\Tau_0(\Phi_2) > \Tau_0^\cut\right) \, ,
\end{align}
where we use the symbol $\overline{\Theta}_{\mathrm{map}}^{\Tau_0}(\Phi_2)$ to denote the region of phase space which cannot be projected on the physical $\Phi_1$ phase space via the $\Phi_1^{\Tau_0} (\Phi_2)$ mapping.
In principle in the $\Tau_0 < \Tau_0^{\rm cut}$ region also events with two hard emissions contribute at NNLO. However, we stress again that when the \zerojet resolution variable is $\pt$ and one keeps $\pt^{\rm cut}$ small enough, these nonsingular contributions from events with two hard jets are negligible and we therefore set
\begin{align}
\frac{\dsigMC_{\geq 2}}{\df\Phi_{2}} (\pt < \pt^{\rm cut}) \equiv 0\,.
\end{align}
This concludes our review of the \geneva method.

In the following we will use the transverse momentum as \zerojet resolution variables and 1-jettiness as \onejet resolution variable.
We shall use the notation \genevapt to label our predictions, while we will use the notation \genevatau when referring to the original \geneva implementation using beam-thrust.
In the next section we discuss the resummation of the transverse momentum distribution in colour-singlet production within the \radish framework.
We refer the reader to \citeRefs{Alioli:2016wqt,Alioli:2019qzz} for the discussion of the NLL resummation of $\mathcal T_1$ within the \geneva formalism.

\subsection{Transverse momentum resummation in the R{\scriptsize AD}ISH formalism}\label{sec:resumm-radish}

Various formalisms to perform the resummation of the transverse momentum in colour-singlet processes have been developed over the last four decades~\cite{Parisi:1979se,Collins:1984kg,Balazs:1995nz,Ellis:1997sc,Balazs:1997xd,Idilbi:2005er,Bozzi:2005wk,Catani:2010pd,Becher:2010tm,Becher:2011xn,GarciaEchevarria:2011rb,Chiu:2012ir,Neill:2015roa,Monni:2016ktx,Ebert:2016gcn,Bizon:2017rah}.
All the ingredients for the N$^3$LL $\pt$ resummation have been computed in \citeRefs{Catani:2011kr,Catani:2012qa,Gehrmann:2014yya,Luebbert:2016itl,Echevarria:2016scs,Li:2016ctv,Vladimirov:2016dll,Moch:2018wjh,Lee:2019zop}, and state-of-the-art predictions for Drell-Yan production now reach this accuracy~\cite{Bizon:2017rah,Bizon:2018foh,Bizon:2019zgf,Bertone:2019nxa,Bacchetta:2019sam,Ebert:2020dfc,Becher:2020ugp}.
In this section we summarise the \radish resummation formalism developed in \citeRefs{Monni:2016ktx,Bizon:2017rah}, which we use in this work to resum the transverse momentum spectrum at N$^3$LL.

The \radish formalism allows one to resum any transverse observable (i.e.~not depending on the rapidity of the
radiation) which fulfils recursive infrared collinear
(rIRC) safety~\cite{Banfi:2004yd}.
The resummation is formulated directly in momentum space by exploiting factorisation properties of squared QCD amplitudes.
The resummation is then evaluated numerically via MC methods.
The RadISH formul\ae\ are more conveniently expressed at the level of the cumulative cross section
\begin{equation}
\label{eq:cumulative}
\Sigma(\pt^{\rm cut}) \equiv \int_0^{\pt^{\rm cut}} \rd \pt\; \frac{\rd \sigma(\pt)}{\rd \pt} = \int \rd \Phi_0  \frac{\df \sigma^{\rm res}}{\df\Phi_{0}} (\pt^{\rm cut}) \,,
\end{equation}
where $\pt=\hat q_\perp(\Phi_0,k_1,\dots,k_n)$ is a function of the Born phase space~$\Phi_0$ of the produced colour singlet
and of the momenta $k_1,...,k_n$ of $n$ real emissions.

For example, in the soft limit, the all-order structure of $\Sigma(\pt^{\rm cut})$, fully differential in the Born phase space, can be written as
\begin{align}
  \label{eq:sigma-2}
  \frac{\df \sigma^{\rm res}}{\df\Phi_{0}} (\pt^{\rm cut}) = &  {\cal V}(\Phi_0) \sum_{n=0}^{\infty}
  \int\prod_{i=1}^n [\rd k_i]
  |{\cal M}(\Phi_0,k_1,\dots ,k_n)|^2\,\nn\\&\quad \times \theta\left(\pt^{\rm cut}-\hat q_\perp(\Phi_0,k_1,\dots,k_n)\right)
\end{align}
Here $[\rd k_i]$  and $\rd\Phi_0$ denote the phase spaces of the $i$-th emission $k_i$ and of the Born configuration, respectively, ${\cal M}$ denotes the matrix element for $n$ real emissions in the soft approximation, and ${\cal V}(\Phi_0)$ is the resummed form factor in the soft limit, which encodes the purely virtual corrections, see~\cite{Magnea:2000ss,Laenen:2000ij,Dixon:2008gr}.
To obtain the resummation it is first necessary to establish a well-defined logarithmic counting in the squared amplitude.
The counting is established by decomposing the squared amplitude defined in \Eq{sigma-2} in $n$-particle-correlated blocks, which contain the correlated portion of the squared $n$-emission soft amplitude and its virtual corrections~\cite{Banfi:2004yd,Banfi:2014sua,Bizon:2017rah}.
In particular, blocks with $n$ particles start contributing one logarithmic order higher than blocks with $(n-1)$ particles, which allows one to systematically identify all the relevant contributions entering at a given logarithmic order.

Thanks to the rIRC safety of the observables, the divergences appearing at all perturbative orders in the real matrix elements cancel exactly those of virtual origin, which are contained in the $\mathcal V(\Phi_0) $ factor of \Eq{sigma-2}.
The cancellation of the singularities is achieved by introducing a resolution scale $q_0$.
Radiation softer than $q_0$ is dubbed unresolved and can be exponentiated to cancel the divergences of virtual origin.
Radiation harder than $q_0$ (\textit{resolved}) must instead be
generated exclusively as it is constrained by the measurement function $\theta\left(\pt^{\rm cut}-\hat q_\perp(\Phi_0,k_1,\dots,k_n)\right)$ in \Eq{sigma-2}.
Recursive IRC safety ensures that neglecting radiation softer than $q_0$ (\textit{unresolved}) in the computation of $\hat q_\perp(\Phi_0,k_1,\dots,k_n)$ only produces terms suppressed by powers of $q_0$, thus ensuring that the limit $q_0\to0$ can be taken safely.

In the \radish formalism the resolution scale $q_0$ is set to a small fraction $\eps > 0$ of the transverse momentum of the correlated block with the largest $k_\perp$, which we henceforth denote as $k_{\perp,1}$.
The cumulative cross section at N$^3$LL accuracy for the production of a colour singlet of mass $M$, fully differential in the Born kinematic variables and including also the effect of collinear radiation can then be written as~\cite{Bizon:2017rah}
\begin{widetext}
\begin{align}
\label{eq:master-kt-space}
\frac{\df \sigma^{\rm res}}{\df\Phi_{0}} (\pt^{\rm cut}) = &  \int\frac{\rd k_{\perp,1}}{k_{\perp,1}}\frac{\rd
  \phi_1}{2\pi}\partial_{\tildeL}\left(-\re^{-\tildeR(k_{\perp,1})} {\tildecalL}_{\rm
  N^3LL}(k_{\perp,1}) \right) \int \dZ\Theta\! \left(\pt^{\rm cut}-  |\vec{k}_{\perp,1}+\dots+\vec{k}_{\perp,n+1}|\right)
                             \notag\\  \notag \\
& + \int\frac{\rd k_{\perp,1}}{k_{\perp,1}}\frac{\rd
  \phi_1}{2\pi} \re^{-\tildeR(k_{\perp,1})} \!  \int \dZ\int_{0}^{1}\frac{\rd \zeta_{s}}{\zeta_{s}}\frac{\rd
  \phi_s}{2\pi}\Bigg\{\bigg({\tildeR}' (k_{\perp,1}) {\tildecalL}_{\rm
  NNLL}(k_{\perp,1}) - \partial_{\tildeL} {\tildecalL}_{\rm
  NNLL}(k_{\perp,1})\bigg) \times\notag \\
&\left(\! {\tildeR}'' (k_{\perp,1}) \!  \ln \! \frac{1}{\zeta_s}\!  + \! \frac{1}{2} {\tildeR}'''
  (k_{\perp,1})\ln^2 \! \frac{1}{\zeta_s} \!  \right)\! -\!  {\tildeR}' (k_{\perp,1})\! \left( \! \partial_{\tildeL} {\tildecalL}_{\rm
  NNLL}(k_{\perp,1})\!  - \!  2\frac{\beta_0}{\pi}\as^2(k_{\perp,1}) \hat{P}^{(0)}\!  \otimes \! {\tildecalL}_{\rm
  NLL}(k_{\perp,1})\!  \ln \! \frac{1}{\zeta_s} \!
\right)\notag\\
&+\frac{\as^2(k_{\perp,1}) }{\pi^2}\hat{P}^{(0)}\otimes \hat{P}^{(0)}\otimes {\tildecalL}_{\rm
  NLL}(k_{\perp,1})\Bigg\} \bigg\{\Theta\left(\pt^{\rm cut}-|\vec{k}_{\perp,1}+\dots+\vec{k}_{\perp,n+1}+\vec{k}_{\perp,s}|\right)\notag  \\
  & - \Theta\left(\pt^{\rm cut}-|\vec{k}_{\perp,1}+\dots+\vec{k}_{\perp,n+1}|\right)\bigg\}\notag\\ \notag \\
& + \frac{1}{2}\int\frac{\rd k_{\perp,1}}{k_{\perp,1}}\frac{\rd
  \phi_1}{2\pi} \re^{-\tildeR(k_{\perp,1})} \int \dZ\int_{0}^{1}\frac{\rd \zeta_{s1}}{\zeta_{s1}}\frac{\rd
  \phi_{s1}}{2\pi}\int_{0}^{1}\frac{\rd \zeta_{s2}}{\zeta_{s2}}\frac{\rd
  \phi_{s2}}{2\pi} {\tildeR}' (k_{\perp,1})\\
&\times\Bigg\{ {\tildecalL}_{\rm
  NLL}(k_{\perp,1}) \left({\tildeR}'' (k_{\perp,1})\right)^2\ln\frac{1}{\zeta_{s1}} \ln\frac{1}{\zeta_{s2}} - \partial_{\tildeL} {\tildecalL}_{\rm
  NLL}(k_{\perp,1}) {\tildeR}'' (k_{\perp,1})\bigg(\ln\frac{1}{\zeta_{s1}}
  +\ln\frac{1}{\zeta_{s2}} \bigg)\notag\\
&+ \frac{\as^2(k_{\perp,1}) }{\pi^2}\hat{P}^{(0)}\otimes \hat{P}^{(0)}\otimes {\tildecalL}_{\rm
  NLL}(k_{\perp,1})\Bigg\}\notag\\
&\times \bigg\{\Theta\left(\pt^{\rm cut}-|\vec{k}_{\perp,1}+\dots+\vec{k}_{\perp,n+1}+\vec{k}_{\perp,s1}+\vec{k}_{\perp,s2}|\right) - \Theta\left(\pt^{\rm cut}-|\vec{k}_{\perp,1}+\dots+\vec{k}_{\perp,n+1}+\vec{k}_{\perp,s1}|\right) -\notag\\ &\Theta\left(\pt^{\rm cut}-|\vec{k}_{\perp,1}+\dots+\vec{k}_{\perp,n+1}+\vec{k}_{\perp,s2}|\right) + \Theta\left(\pt^{\rm cut}-|\vec{k}_{\perp,1}+\dots+\vec{k}_{\perp,n+1}|\right)\bigg\}\!.\notag
\end{align}
\end{widetext}

In the equation above, the first line enters already at NLL, the first set of curly brackets
(second to fifth line)
starts contributing at NNLL, and the last set of curly brackets
(from line six) enters at N$^3$LL.

The functions $\tildecalL$ are luminosity factors evaluated at different orders which involve, besides the parton luminosities, the process-dependent squared Born amplitude and hard-virtual corrections $\tildeH^{(n)}$ as well as the coefficient functions $\tildeC_{ci}^{(n)}$, which have been evaluated to second order for $q \bar q$-initiated processes in \citeRefs{Catani:2012qa,Gehrmann:2014yya,Echevarria:2016scs}.
The factors $\hat{P}^{(0)}$ denote the regularised splitting functions.
The interested reader is referred to section 4 of \citeRef{Bizon:2017rah} for the definition of the luminosity factors and their ingredients.
We defined $\zeta_{si} \equiv k_{\perp,si}/k_{\perp,1}$ and we introduced the notation $\dZ$ to denote an ensemble describing the emission of $n$ identical independent blocks.
The average of a function $G(\Phi_0,\{k_i\})$ over the measure
$\rd {\cal Z}$  as ($\zeta_{i} \equiv k_{\perp,i}/k_{\perp,1}$) is defined as
\begin{align}
\label{eq:dZ}
\int \! &    \dZ  G(\Phi_0,\{k_i\})\! =\! \re^{-{\tildeR}'(k_{\perp,1})\ln\frac{1}{\eps}}\ \times\\&\quad
   \sum_{n=1}^{\infty}\!  \frac{1}{n!} \! \prod_{i=2}^{n+1} \!
    \int_{\eps}^{1} \!  \! \frac{\rd\zeta_i}{\zeta_i} \!  \!  \int_0^{2\pi}\!  \!
   \frac{\rd\phi_i}{2\pi} \!  {\tildeR}'(k_{\perp,1})G(\Phi_0,k_1,\dots,k_{n+1})\, \nn.
\end{align}
Note that the $\ln 1/\eps$ divergence appearing in the exponential prefactor of \Eq{dZ} cancels exactly that in the resolved real radiation, encoded in the nested sums of products on the right-hand side of \Eq{dZ} .

\Eq{master-kt-space} has been obtained by expanding all the ingredients around $k_{\perp,1}$ since $\zeta_i =  k_{\perp,i}/k_{\perp,1} \sim \mathcal O(1)$, see \citeRef{Bizon:2017rah}.
Thanks to rIRC safety, blocks with $k_{\perp,i} \ll k_{\perp,1}$ are fully cancelled by the term $\exp\{-{\tildeR}'(k_{\perp,1})\ln(1/\eps)\}$ of \Eq{dZ}.
Although such an expansion is not strictly necessary, it makes a numerical implementation much more efficient~\cite{Bizon:2017rah}.
Because of this expansion, \Eq{master-kt-space} contains explicitly the derivatives
\begin{equation}\label{eq:Rderivs}
{\tildeR}'= \rd \tildeR/\rd\tildeL\,,\quad {\tildeR}''= \rd
      {\tildeR}'/\rd\tildeL\,,\quad {\tildeR}'''= \rd {\tildeR}''/\rd\tildeL
\end{equation}
of the radiator $\tildeR$, which is given by
\begin{align}
\label{eq:mod-radiator}
\tildeR(k_{\perp,1}) = &- \tildeL g_1(\as \beta_0\tildeL ) -
  g_2(\as \beta_0\tildeL )  \nn\\
  &- \frac{\as}{\pi} g_3(\as \beta_0\tildeL ) - \frac{\as^2}{\pi^2}
  g_4(\as \beta_0\tildeL )\,,
\end{align}
with $\as = \as(\mu_R)$ and $\mu_R\sim M$ being the renormalisation scale, where $M$ is the hard scale of the process.
Here $L=\ln({Q}/{k_{\perp,1}})$, where $Q\sim M$ is the resummation scale, whose variation is used to probe the size of missing logarithmic higher-order corrections in \Eq{master-kt-space}.
The functions $g_i$ are reported in Eqs.~(B.8-B.11) of \citeRef{Bizon:2017rah}.

The above formul\ae\ are implemented in the code \radish, which evaluates them using MC methods.
We refer the reader to section~4.3 of \citeRef{Bizon:2017rah} for the technical details.

The resummed result provided by \radish is valid in the soft/collinear region, i.e.~$\pt/M \ll 1$, and must be matched with fixed-order predictions at large values of $\pt$.

Resummation effects should thus vanish in the large $\pt$ region.
This is enforced by mapping the limit $\kto \rightarrow Q$, where the logarithms vanish, onto $\kto \rightarrow \infty$ via modified logarithms
\begin{equation}\label{eq:modlog}
	\ln \frac{Q}{\kto} \rightarrow \tilde{L} \equiv\frac{1}{p} \ln \left( \left(\frac{Q}{\kto}\right)^p + 1 \right),
\end{equation}
where $p$ is a positive real parameter whose value is chosen such that the resummed component decreases faster than the fixed-order spectrum for $\pt/M \gtrsim 1$.
In the following, we took $p=4$.
Therefore, the logarithms $L$ in the Sudakov radiator~\Eq{mod-radiator}, its derivatives \Eq{Rderivs} and the luminosity factors have to be replaced by $\tilde L$.

For consistency, one must supplement \Eq{master-kt-space} with the following Jacobian:
\begin{equation}
	\mathcal{J} (\kto) =  \left(\frac{Q}{\kto}\right)^p \left( \left(\frac{Q}{\kto}\right)^p + 1 \right)^{-1}.
\end{equation}
This prescription leaves the measurement functions in \Eq{master-kt-space} unchanged.
The final result is modified beyond the nominal accuracy by power
corrections in $(Q/\kto)^p$.

\section{Implementation details}\label{sec:implementation}

In this section we discuss the interface of the \radish code with \geneva and the
 implementation of the Drell-Yan process in \geneva using $\pt$ as \zerojet resolution variable.
We validate our framework by comparing our results with NNLO predictions obtained with \Matrix~\cite{Grazzini:2017mhc} and discuss the interface with parton showers.

\subsection{Interfacing G{\scriptsize ENEVA} with R{\scriptsize AD}ISH}\label{sec:radishimpl}

The resummation formul\ae\ discussed in Sect.~\ref{sec:resumm-radish} are implemented in the \texttt{fortran90} code \radish.
 For each Born event the code produces the initial-state radiation and performs the resummation of large logarithmic contributions using MC methods as described in Sect.~4.3 of \citeRef{Bizon:2017rah}.
As a result, the cumulative distribution is filled on-the-fly yielding $\df \sigma^{\rm res}/ \df\Phi_{0} \,  (\pt^{\rm cut})$.
The spectrum can be calculated by differentiating the cumulative histogram numerically.

We note that our need to obtain the spectrum at a given value of $\pt$ as in \Eq{1incmaster} poses some technical challenges which need to be addressed in order to interface \radish with \geneva.
Indeed, the \radish code has been designed to compute the whole cumulant distribution by generating an MC ensemble of soft and/or collinear emissions.
The simplest solution would be to run the MC algorithm with a very large number of events for each Born configuration provided by \geneva in order to yield sufficiently stable results for the numerical derivative of the cumulant.
However, despite the effectiveness of the MC generation, this on-the-fly computation is rather inefficient.

A better approach is to first compute the \radish cumulant with a very large number of Born points, exploiting runtime parallelisation. This is done before starting the main \geneva runs. From these parallel \radish runs we build an interpolation grid, which  is then used to provide the cumulant and the spectrum for the \geneva runs on an event-by-event basis.
To this end, we first parametrise the Born phase space using two variables\footnote{Specifically, we use the rapidity of the virtual boson $Y_{\ell \ell}$ and $a_{\ell\ell} = \arctan((m_{\ell\ell}^2 -m_Z^2)/(m_Z \Gamma_Z))$ for Drell-Yan production. The latter variable is chosen to flatten the $m_{\ell\ell}$ distribution, see e.g.~\cite{Astill:2016hpa}.} and construct a discrete lattice in these two parameters.
We then compute the resummed cumulant in each bin and for each combination of perturbative scales up to the maximal value of $\pt$ kinematically allowed.
The resulting grids are then loaded and interpolated on-the-fly by means of Chebyshev polynomials as implemented in the \texttt{GSL} library~\cite{GSL2009}, yielding a fast evaluation of the cumulant and of its derivative.
Since the shape of the resummed result depends only mildly on the Born variables, it is sufficient to rescale the resummation by the Born squared amplitude to obtain a result differential in the Born kinematics.
We found that a $10 \times 10$ dimensional grid in the Born variables provides a sufficiently fine discretisation of the Born phase space.

The other ingredient which is provided by \radish is the expansion of the resummation to remove the double counting between resummation and fixed order.
In this case the use of interpolation routines is not advisable, since one must ensure the exact cancellation of the large logarithmic terms at low $\pt$ to avoid spurious and potentially large effects on the final results.\footnote{Indeed, in a first attempt using an interpolation also for the resummed expanded contribution, we observed numerical instabilities in the $Y_{\ell\ell}$ distribution.}
However, since the expansion is computed semi-analytically in \radish (for additional details we refer the reader to Sect.~4.2 of \cite{Bizon:2017rah}) we can avoid the interpolation altogether and compute it on-the-fly without affecting the speed of the code.

\begin{figure}[t]
  \includegraphics[width=\rescaleoneplot]{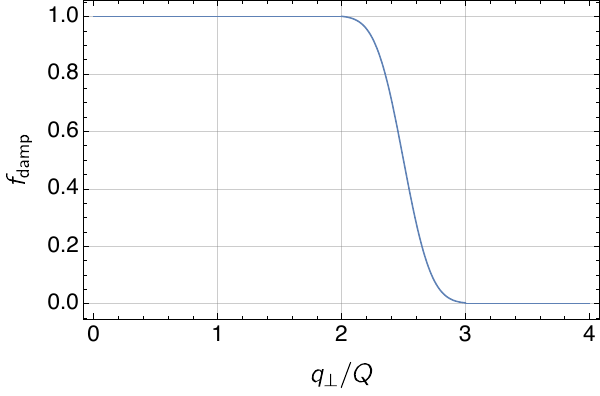}%
  \caption{Damping function used to further suppress the resummation at large values of $\pt$, see text for details.}
  \label{fig:damping}
\end{figure}

Once the values of the resummation and the expansion are available, the matched results for the spectrum of \Eq{1incmaster} can be obtained by means of a standard additive matching.
The use of modified logarithms~\eqref{eq:modlog} automatically ensures that the effect of the resummation vanishes in the fixed-order region.
However, numerical instabilities could arise since, while the expansion is exact, the resummed result is obtained by means of a Chebyshev interpolation. This can induce tiny but visible numerical artefacts in the region where the cumulant flattens and its derivative approaches zero.
In order to avoid this unwanted feature, we introduce a function which further suppresses both the resummed and the resummed expanded results smoothly in the fixed-order region, where a complete cancellation between them should be achieved.
In this way, undesired numerical instabilities in this region are removed.
In particular, we use the following function:
\begin{equation}
	f_{\rm damp} (\pt) =
	\begin{cases}
	1 & \pt < l\\
	\frac{1 - {\rm erf}\left( \frac{2( \pt - m)}{w} \right)}{ 1 - {\rm erf}\left( \frac{2 (l - m )}{w}\right)}	& l < \pt < r\\
	0 & \pt > r
	\end{cases}
\end{equation}
with $   m = (r+l)/2$, $  w = (r-l)/2$ and
where we take $l = 2 Q$ and  $r = 3 Q$ and `erf' is the standard Gaussian error function.
We observe that with this choice there is a very tiny discontinuity at $\pt = r$, which however does not give any visible effect since $f_{\rm damp}(r-\epsilon) \simeq 0$.
The damping function is plotted as a function of $\pt/Q$ in \Fig{damping}.

\subsection{Power-suppressed corrections to the nonsingular cumulant}\label{sec:power-suppressed-terms}

In \Eq{0full} we wrote the expression for the NNLO accurate \zerojet
cross section fully differential in the Born phase space, whose implementation would require a local NNLO subtraction scheme.
In our implementation the NNLO$_0$ accuracy is instead achieved replacing \Eq{0full} with the following expression 
\begin{align}
\frac{\widetilde\dsigMC_0}{\df\Phi_0}&(\Tau_0^\cut) =
\frac{\df\sigma^{\rm res}}{\df\Phi_0}(\Tau_0^\cut)\, - \biggl[\frac{\df\sigma^{\rm res}}{\df\Phi_{0}}(\Tau_0^\cut) \biggr]_{\rm NLO_0} \, \nn\\
&+(B_0+V_0)(\Phi_0)  \nn \\
&+  \int \frac{\mathrm{d} \Phi_1}{\mathrm{d} \Phi_0} (B_1)(\Phi_1)\,\theta\big( r_0(\Phi_1)< \Tau_0^{\mathrm{cut}}\big),
\label{eq:0widetilde}
\end{align}
which only involves a local subtraction at $\mathcal O(\as) $ and the expansion of the resummation at the same order.
The formula assumes an exact cancellation between the fixed-order and the resummed expanded contribution below the value of $\Tau_0^{\rm cut}$ at order $\as^2$.
The cancellation is guaranteed for the singular contributions due to the accuracy of the $\Tau_0$ cumulant (we stress again that in order to achieve NNLO$_0$ accuracy the resummation accuracy must be at least NNLL$^\prime$); however, the formula fails to capture nonsingular contributions at $\mathcal O(\as^2)$.
These nonsingular contributions can be expressed as
\begin{align} \label{eq:Sigmanons}
 \frac{\df\sigma_0^\nons}{\df\Phi_{0}}(\Tau_0^\cut)
&= \bigl[ \as f_1(\Tau_0^\cut, \Phi_0) + \as^2 f_2(\Tau_0^\cut, \Phi_0) \bigr]\, \Tau_0^\cut \, ,
\end{align}
and  their integral over the phase space is
\begin{align}
\Sigma_{\mathrm{ns}} (\Tau_0^\cut) & = \int \df \Phi_0 \,
\frac{\df\sigma_0^\nons}{\df\Phi_{0}}(\Tau_0^\cut) \, .
\end{align}
The nonsingular cumulant vanishes in the
limit $\Tau_0^\cut\to 0$ since the functions $f_i(\Tau_0^\cut, \Phi_0)$ contain at worst
logarithmic divergences.
As discussed above, our calculation includes the term $f_1(\Tau_0^\cut, \Phi_0)$ since we implement a NLO$_1$ FKS local subtraction.
On the contrary, $f_2(\Tau_0^\cut, \Phi_0)$ is not included in \Eq{Sigmanons}.
Neglecting the $\mathcal O(\as^2)$ power corrections is acceptable as long as we choose
$\Tau_0^\cut$ to be very small.

So far, the \geneva method has been based on $N$-jettiness subtraction~\cite{Gaunt:2015pea,Boughezal:2015dva,Boughezal:2015eha}.
In this work we take $\Tau_0$ equal to $\pt$; effectively, this corresponds to basing \geneva on a $q_T$ subtraction scheme~\cite{Catani:2007vq}.
The availability of different resolution parameters in \geneva is beneficial, as the size and the scaling of the power corrections can be different.
The size of the missing $\mathcal O(\as^2)$  power corrections as a function of $\pt^\cut$ can be calculated as
\begin{align}
\Sigma_{\textrm{NS}}^{(2)}(\pt^\cut) &=  \delta \sigma^{\textrm{NNLO}}  - \Sigma^{\textrm{N$^3$LL}}_{\rm asymp} + \Sigma^{\textrm{N$^3$LL}}_{\rm asymp}|_{\alpha_s} \nonumber \\ & \quad -
\int_{\pt^\cut}^{\infty}  \df \pt \, \left(\frac{\df \sigma^{\textrm{NLO}_1}}{\df \pt} - \left.\frac{\df \sigma^{\textrm{N$^3$LL}}}{\df \pt}\right|_{\alpha^2_s}\right)
 \nonumber \\ & \quad +
\int_{\pt^\cut}^{\infty}  \df \pt \, \left(\frac{\df \sigma^{\textrm{LO}_1}}{\df \pt} - \left.\frac{\df \sigma^{\textrm{N$^3$LL}}}{\df \pt}\right|_{\alpha_s}\right)
\,,
\end{align}
where $\delta \sigma^{\textrm{NNLO}} =  \sigma^{\textrm{NNLO}} - \sigma^{\textrm{NLO}}$, $\Sigma^{\textrm{N$^3$LL}}_{\rm asymp}$ is the cumulant of the resummed contribution (see \Eq{cumulative})
\begin{equation}
	\Sigma^{\textrm{N$^3$LL}}_{\rm asymp} = \int \df \Phi_0  \int_0^{\pt^{\rm max}} \df \pt' \,
\frac{\df\sigma^{\textrm{N$^3$LL}}}{\df\Phi_{0} \df \pt'}
\end{equation}
and $\pt^{\rm max}$ is the maximum value allowed by the kinematics for each $\Phi_0$ point. Finally,  $|_{\as^n}$ indicates the expansion up to order $\as^n$.

We have calculated $\delta \sigma^{\textrm{NNLO}}$ by computing $\sigma^{\textrm{NNLO}} $ with~\Matrix.
Note that \Matrix also achieves NNLO accuracy via $q_T$ subtraction, and therefore potentially misses power corrections at $\mathcal O(\as^2)$.
However, \Matrix includes an estimate of this power corrections by interpolating the result to $\pt^\cut = 0$, and including an estimate of this interpolation procedure in its error.

\begin{figure}[t!]
  \includegraphics[width=\rescaleoneplot]{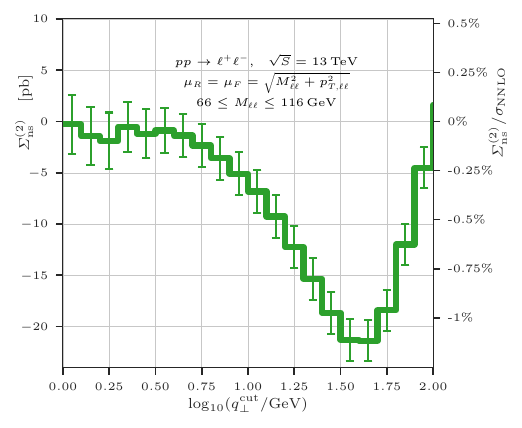}
	\caption{Nonsingular cumulant at order $\as^2$ as a function of $\pt^{\rm cut}$.}
	\label{fig:nscumulant}
\end{figure}

We show the size of the missing power corrections in~\Fig{nscumulant}, where we consider values of $\pt^\cut$ down to $1$~GeV.
We consider $pp \rightarrow \ell^+ \ell^-$ production at $13$ TeV in an inclusive setup by applying a cut only on the invariant mass of the produced colour singlet.
We observe that the power corrections are below the $2 \permil $ level for $\log_{10} (\pt^\cut) \lesssim 0.5$, corresponding to a value of $\pt^\cut \lesssim 3$~GeV, in accordance with what was observed in \cite{Grazzini:2017mhc}.
Motivated by this plot, we choose $\pt^\cut=1$~GeV as our default value.
The negligible size of the missing power correction allows us to avoid the need for the reweighting of the $\Phi_0$ events, which was instead the approach followed in the previous application of the \geneva method.
Our choice is further justified by a detailed comparison between \geneva and an independent NNLO calculation for distributions differential in the $\Phi_0$ variables, as we will show in \Sec{nnlocomp}.

\begin{figure*}[t!]
  \begin{subfigure}[b]{\rescaletwoplots}
    \includegraphics[width=\textwidth]{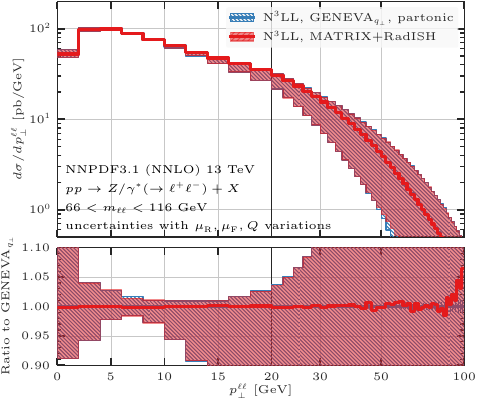}%
  \end{subfigure}
  \hspace*{\hspacebetweentwoplots}
  \begin{subfigure}[b]{\rescaletwoplots}
    \includegraphics[width=\textwidth]{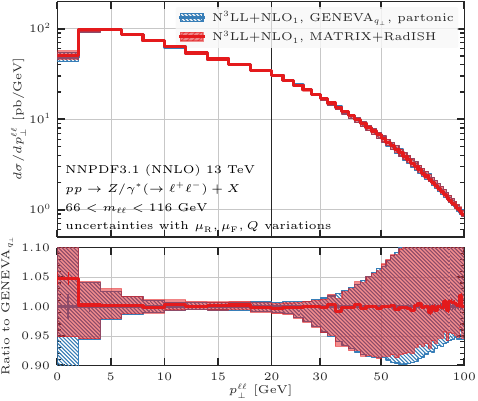}%
  \end{subfigure}
  \vspace{\spacebeforefigurecaption}
  \caption{Comparison between the resummed N$^3$LL (left panel) and matched N$^3$LL+NLO$_1$ (right panel) results obtained with \textsc{Matrix}+\radish and with \mbox{\geneva{}+\radish}.}
	\label{fig:n3llcmp}
\end{figure*}

\subsection{NLO$_1$ calculation and phase space mapping}

In this section we discuss the implementation of the $\Phi_1^r(\Phi_2)$ mapping introduced in \Eq{1masterfull} and of that introduced in \Eq{P2norm}.
As we anticipated in \Sec{geneva}, these phase space mappings do not necessarily need to coincide, since only the latter needs to be written as a function of the \onejet resolution variable $\mathcal T_1$.

We start by discussing the mapping used to implement the NLO$_1$ calculation.
Let us first notice that the $\Phi_1^r(\Phi_2)$ mapping used in the $B_2$ term and the $\Phi_2^C (\Phi_1)$ used for the $C$ term in \Eq{1masterfull} can be different, but provided that both are IR safe, they must be equivalent in the IR singular limit.
However, as we discussed in \Sec{geneva}, the $\Phi_1^r(\Phi_2)$ mapping must have the additional property that it preserves the \zerojet resolution variable $\pt$, in order to guarantee that the NLO$_1$ is pointwise consistent with the $\pt$ resummation.
Moreover, the mapping must be invertible, meaning that one should be able to reconstruct all the $\Phi_2$ points which can be projected on a given $\Phi_1$ configuration.
The additional requirement that the mapping should preserve $\pt$ poses some challenges as
discussed in Sec.~A6 of \citeRef{Alioli:2015toa}.
In addition to preserving the 0-jettiness $\mathcal T_0$ which was used as \zerojet resolution variable\footnote{More precisely, the mapping preserves a recursive definition of $\mathcal T_0$ dubbed $\mathcal T_0^{\rm FR}$, see \citeRef{Alioli:2015toa} for additional details.} the  $\mathcal T_0^{\rm FR}$ mapping introduced in \citeRef{Alioli:2015toa} also preserves the four momentum of the vector boson $q^\mu (\Phi_2) = q^\mu (\Phi_1^{\mathcal T} (\Phi_2)) $ and thus the full transverse momentum $\vec q_\perp$.
Therefore, in our study one could in principle use the same mapping to implement the NLO$_1$ calculation and the splitting function $\mathcal P(\Phi_2)$ used in the $\mathcal T_1$ resummation.

It happens, however, that the $\mathcal T_0^{\rm FR}$ mapping creates undesired higher-order artefacts in a few NLO$_1$ differential distributions when used also in the $\mathcal T_1$ resummation. In particular, one observes effects on quantities such as the rapidity separation between the leading jet and the vector boson.
For this reason, in all recent applications of the \geneva method based on the resummation of $\mathcal T_0$, the $\mathcal T_0^{\rm FR}$ mapping has been modified, relaxing the condition that it preserves the transverse momentum of the vector boson.
Therefore, while one could use the $\vec q_\perp$-preserving $\mathcal T_0^{\rm FR}$ mapping for the subtraction, one needs a new phase-space mapping which preserves $\vec q_\perp$ for the $\mathcal T_1$ resummation projection.
For the initial state radiation (ISR), we found that the $q^\mu$-preserving mapping introduced in \citeRef{Figy:2018imt} does not create distortions in any differential distribution, and we have therefore implemented this mapping in the \geneva code.
For final-state radiation (FSR), a mapping preserving the four-momentum $q^\mu$ of the vector boson is uniquely defined once one decides how to map the (massive) jet with momentum $p_{12}^\mu = p_1^\mu + p_2^\mu$ of the 2-jet configuration into a massless jet with momentum $\bar p^\mu$ in the \onejet configuration.
In particular, we choose to conserve the longitudinal component of the jet, i.e.~$ p_1^z + p_2^z=\bar p^z$, as described in Appendix \ref{sec:mapping}.

We have verified that these mappings can be used both in the NLO$_1$ calculation (in \Eq{1masterfull}) as well as in the $\mathcal T_1$ resummation (in \Eq{2masterful}).
We have however preferred to maintain the ($\vec q_\perp$-preserving) $\mathcal T_0^{\rm FR}$ mapping in the NLO$_1$ calculation, due to slightly better convergence properties, while we resort to the new ISR and FSR mappings in the implementation of the $\mathcal T_1$ resummation. In this way we avoid unnecessarily large higher-order effects in the distributions.

\subsection{Validation of the N$^3$LL resummation}\label{sec:val1}

In this section we validate the implementation of the N$^3$LL resummation in \geneva by comparing the matched $\pt$ spectrum with an independent calculation obtained using the \Matrix{}+\radish interface~\cite{Kallweit:2020gva}.
We note that the approach taken in the \geneva framework is somewhat different from that taken in \Matrix{}+\radish.
In the \geneva approach each event generated is assigned a resummed weight according to the value of the chosen resolution parameter; as a consequence, the information about the physical event, which is characterised by a particular value of $\pt$, is always retained throughout the calculation.
As shown in \cite{Ebert:2020dfc}, by retaining the exact lepton kinematics the \genevapt result effectively resums also the linear power corrections which appear as soon as one imposes fiducial cuts~\cite{Ebert:2019zkb}, beyond the $\mathcal O(\as^2)$, which is in any case included after the matching to NNLO.
This is not the case in the \Matrix{}+\radish calculation, where the matching is instead performed at the level of the final distribution and therefore the linear power corrections associated to fiducial cuts are included only up to order $\mathcal O(\as^2)$.\footnote{Note that this is not an intrinsic limitation, since by implementing an appropriate recoil scheme directly in \radish one could achieve the same result.}

The predictions for the $\pt$ spectrum obtained in the two approaches are equivalent in the absence of fiducial cuts on the Born-level variables (see e.g.~the discussion in Sec. 4.3 of \cite{Alioli:2020qrd}).
In the presence of cuts, the two approaches are equivalent only in the limit $\pt \rightarrow 0$, unless the quantity subject to the cuts is preserved by the phase-space mapping $\Phi_N^r (\Phi_{N+1})$.
For the following validation, the only process-defining cut is applied to the invariant mass of the resulting colour singlet, which is preserved by our mappings.
Therefore, we expect full agreement above $\pt^{\rm cut}$ between our results and those obtained with \Matrix{}+\radish, which we can use to validate our implementation.

To produce our predictions we use the NNLO {\tt NNPDF3.1} parton distribution function
set~\cite{Ball:2017nwa} with $\as(M_Z) = 0.118$ through the \textsc{Lhapdf} interface~\cite{Buckley:2014ana}.
We set the central renormalisation and factorisation scales to $m_T = \sqrt{m_{\ell \ell}^2 + 
\pt^2}$, while the central resummation scale is set to $Q=m_{\ell \ell}/2$.
The uncertainty band is constructed as the envelope of a canonical $7$-point variation of the renormalisation ($\mu_{\rm R}$) and factorisation ($\mu_{\rm F}$) scales and of two additional variations of $Q$ by a factor of two for central $\mu_{\rm R}$ and $\mu_{\rm F}$. 
The comparison is presented in \Fig{n3llcmp}, where we show the resummed and the matched results for the $\pt \equiv p_\perp^{\ell \ell} $ spectrum up to $100$~GeV, on a scale which is linear up to $20$~GeV and logarithmic for larger values.
In both cases we observe a very good agreement between the two calculations, both for the central prediction and for the theory uncertainty bands.
We observe marginal deviations at large $\pt$ in the resummed result, which can be traced back to the additional damping present in the \geneva implementation as discussed in \Sec{radishimpl}.
As expected, this difference cancels in the matched result.
The difference in the first bin of the matched result, on the other hand, originates from the sensitivity to the different IR cutoffs employed in the two calculations, necessary to regularise the $\pt \rightarrow 0$ limit in the NLO$_1$ calculation.
We observe that after matching the uncertainty bands are significantly reduced between $10$ and $25$~GeV and are at the 1-2\% level, which might not reflect the actual size of missing higher-order effects. However, since we are not including  further sources of uncertainty such as PDF errors, mass effects, etc.  one should not regard this as the full theoretical uncertainty. Including all these additional effects in order to provide a fully realistic error estimate is beyond the scope of this work.

\begin{figure*}[t]
  \begin{subfigure}[b]{\rescaletwoplots}
    \includegraphics[width=\textwidth]{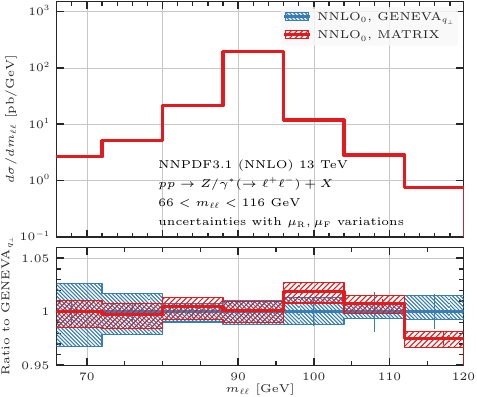}%
  \end{subfigure}
  \hspace*{\hspacebetweentwoplots}
  \begin{subfigure}[b]{\rescaletwoplots}
    \includegraphics[width=\textwidth]{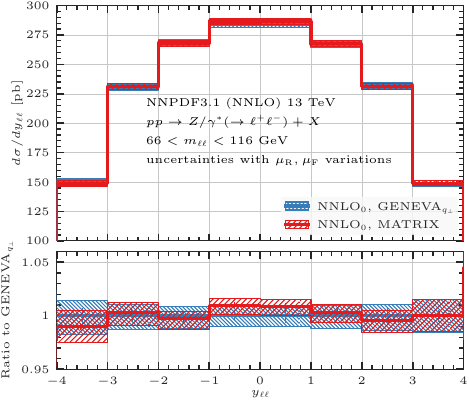}%
  \end{subfigure}
  \begin{subfigure}[b]{\rescaletwoplots}
    \includegraphics[width=\textwidth]{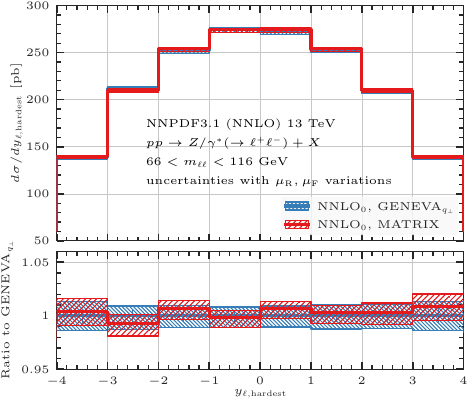}%
  \end{subfigure}
  \hspace*{\hspacebetweentwoplots}
  \begin{subfigure}[b]{\rescaletwoplots}
    \includegraphics[width=\textwidth]{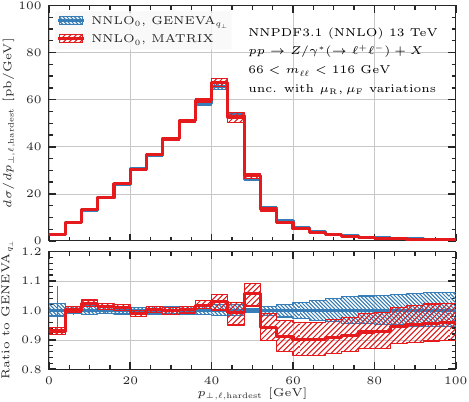}%
  \end{subfigure}
  \vspace{\spacebeforefigurecaption}
  \caption{Comparison between the NNLO results obtained with \textsc{Matrix} and the results obtained with \geneva{}+\radish. Upper panel: invariant mass (left) and rapidity (right) of the lepton pair. Lower panel: rapidity (left) and transverse momentum (right) of the hardest lepton.}
	\label{fig:nnlocmp}
\end{figure*}

\subsection{Comparison with NNLO predictions and validation}\label{sec:nnlocomp}

In this section we proceed with the validation of the \geneva implementation by comparing our results with those obtained with an independent NNLO calculation using \Matrix.
In~\Fig{nnlocmp} we compare four distributions already defined at the Born level, which allows us to assess the agreement of the NNLO corrections.
In particular, we compare the invariant mass $m_{\ell \ell}$ and the rapidity $y_{\ell \ell}$ of the dilepton system, the rapidity of the hardest lepton $y_{\ell, \rm hardest}$ and the transverse momentum of the hardest lepton $p_{T,\ell,\rm hardest}$.
We use the same settings as specified in the previous section, but we now use a canonical $7$-point scale uncertainty prescription by keeping the resummation scale fixed to its central value $Q$.

For the first three observables we find an excellent agreement between \geneva and \Matrix, with differences at the few percent level, which is within the size of the statistical fluctuations.
In the case of $p_{T,\ell,\rm hardest}$ we expect possible differences because the distribution is NNLO accurate only below the Jacobian peak at $m_{\ell \ell}/2$.
The presence of resummation effects in the \geneva results improve the description of this observable above the peak with respect to the pure fixed-order result, smearing out the unphysical behaviour of the distribution around the Sudakov shoulder~\cite{Catani:1997xc}.
Indeed, we observe that the two predictions are in good agreement up to $\sim 40$ GeV.
Above this value, the two predictions start to differ, and the two uncertainty bands barely overlap between $50$ and $70$ GeV.
At larger values of $p_{T,\ell,\rm hardest}$, the two predictions approach each other again.

In conclusion, the comparison between the \geneva and the \Matrix predictions provides a robust check of our implementation.
In particular, the agreement observed at the differential level fully justifies the value of the $\pt^{\rm cut}$ of $1$~GeV which we shall use to produce our results.

\subsection{Interface with the P{\scriptsize YTHIA}8 parton shower}\label{sec:shower}

In order to make a calculation fully differential at higher
multiplicities the \geneva partonic predictions  must be matched to a parton
shower. 
The shower adds extra radiation to the exclusive 0- and \onejet cross section
and extends the inclusive 2-jet cross section by including higher jet
multiplicities. For the sake of definiteness in this study, as in the previous ones, we focus on the \pythiaEight shower.
The \geneva interface to \pythiaEight for hadronic processes is discussed in detail in \citeRef{Alioli:2015toa}; here we briefly summarise the most relevant features, and we discuss the modifications to the interface needed to accommodate the change in the \zerojet resolution variable.
In the previous \geneva implementations based on $\mathcal T_0$ resummation, the shower
constraints aim at preserving as far as possible the NNLL$^\prime$ accuracy of the $\mathcal T_0$ distribution.
In particular, it was shown that  for the majority of the events the first emission of the shower produces a distortion of the
$\mathcal T_0$ distribution at the $\mathcal O(\as^3/\mathcal T_0)$ level,
i.e.~beyond NNLL$^\prime$~\cite{Alioli:2015toa}.
However, due to the vectorial nature of $\pt$, one can not readily apply the same argument in this case.
In general one expects that any emission of the shower could alter the transverse momentum distribution of the colour-singlet system, and ultimately the logarithmic accuracy for the transverse momentum spectrum after the shower is therefore dictated by the shower accuracy.
Hence, in this work we refrain from making any claims about the formal accuracy of the predictions for the $\pt$ spectrum  after the showering.
However, we will show below that it possible, with a suitable choice of the shower recoil scheme, to obtain an excellent numerical agreement between the analytic N$^3$LL resummation and the \geneva showered results.
Lastly, one can get an excellent description of the data at small $\pt$ by tuning the \pythiaEight nonperturbative parameters.
However, in any calculation obtained by matching higher-order calculations with parton shower one has to carefully evaluate which parameters are truly encoding nonperturbative effects and should therefore be tuned.   

\begin{figure*}[t]
  \begin{subfigure}[b]{\rescaletwoplots}
    \includegraphics[width=\textwidth]{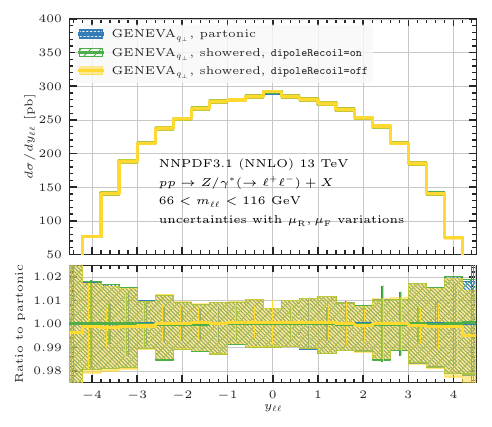}%
  \end{subfigure}
   \hspace*{\hspacebetweentwoplots}
  \begin{subfigure}[b]{\rescaletwoplots}
    \includegraphics[width=\textwidth]{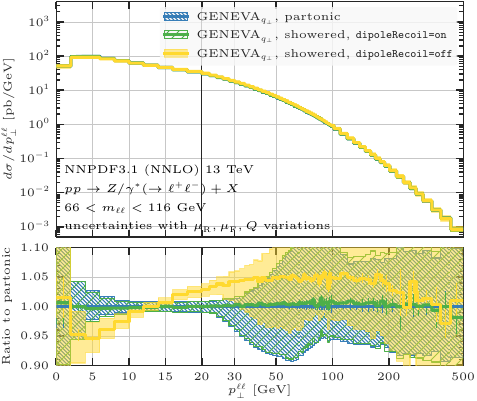}%
  \end{subfigure}
  \vspace{\spacebeforefigurecaption}
  \caption{Comparison between the parton level and the showered results with and without the \pythiaEight \texttt{dipoleRecoil} option for the rapidity distribution (left) and for the transverse momentum distribution (right) of the lepton pair.}
	\label{fig:showercmp}
\end{figure*}

\begin{figure*}[t]
  \begin{subfigure}[b]{\rescaletwoplots}
    \includegraphics[width=\textwidth]{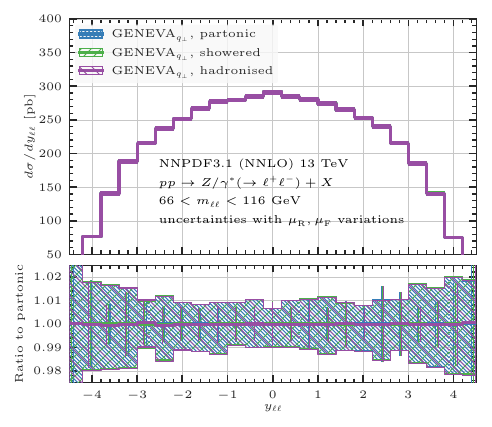}%
  \end{subfigure}
  \hspace*{\hspacebetweentwoplots}
  \begin{subfigure}[b]{\rescaletwoplots}
    \includegraphics[width=\textwidth]{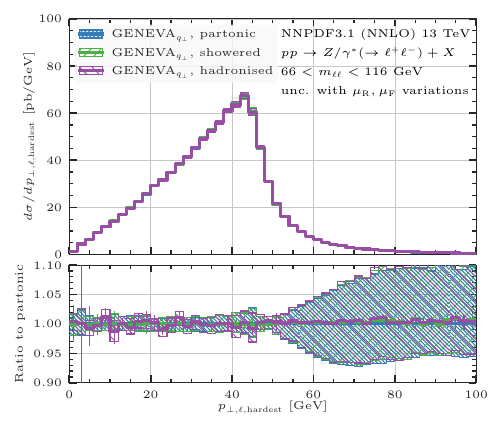}%
  \end{subfigure}
  \vspace{\spacebeforefigurecaption}
  \caption{Comparison of the partonic, showered and hadronised results for the rapidity of the lepton pair (left) and the transverse momentum of the hardest lepton (right).}
	\label{fig:nnlopscmp1}
\end{figure*}

In order to discuss the details of the matching, let us start by analysing the
interface to the \pythiaEight $k_\perp$-ordered parton shower, starting with the
\zerojet event case.
For this set of events, the shower should simply restore the emissions which were integrated over in the construction of the \zerojet cross section.
In our implementation we set the starting scale to these event to the
natural scale $\pt^{\rm cut}$; to avoid double-counting with events
above the cut we require that after the shower the transverse momentum
of the boson does not exceed $\pt^{\rm cut}$, though we allow for a
small spillover if the showered event has $\pt > 1.05 \times \pt^{\rm
  cut}$ to avoid an hard cutoff.
Events which do not fulfill this constraint are re-showered.
In practice, this spillover has a negligible effect, since \zerojet events account for
$\mathcal O(1\%)$ or less of the total cross section, and are therefore a very
small fraction of the total; moreover, since the starting scale for the
showering of these events is $\sim 1 \, \rm GeV $ the majority of them
automatically satisfy the constraint.

\begin{figure*}[t]
  \begin{subfigure}[b]{\rescalethreeplots}
    \includegraphics[width=\textwidth]{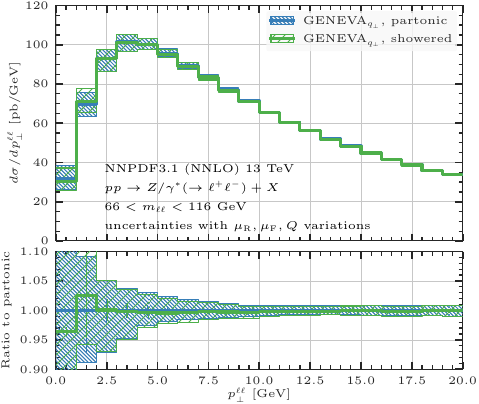}%
  \end{subfigure}
   \hspace*{\hspacebetweenthreeplots}
  \begin{subfigure}[b]{\rescalethreeplots}
    \includegraphics[width=\textwidth]{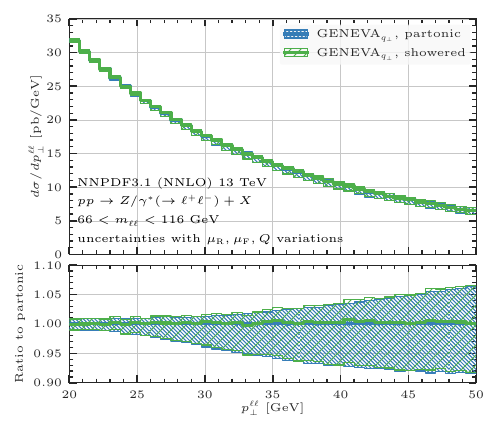}%
  \end{subfigure}
  \hspace*{\hspacebetweenthreeplots}
  \begin{subfigure}[b]{\rescalethreeplots}
    \includegraphics[width=\textwidth]{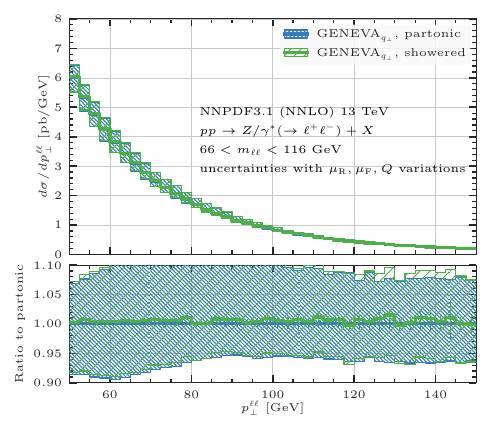}%
  \end{subfigure}\\
  \begin{subfigure}[b]{\rescalethreeplots}
    \includegraphics[width=\textwidth]{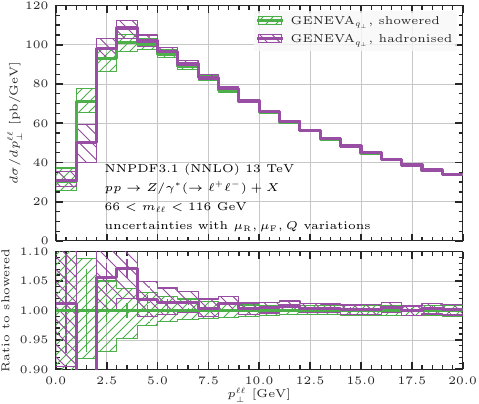}%
  \end{subfigure}
   \hspace*{\hspacebetweenthreeplots}
  \begin{subfigure}[b]{\rescalethreeplots}
    \includegraphics[width=\textwidth]{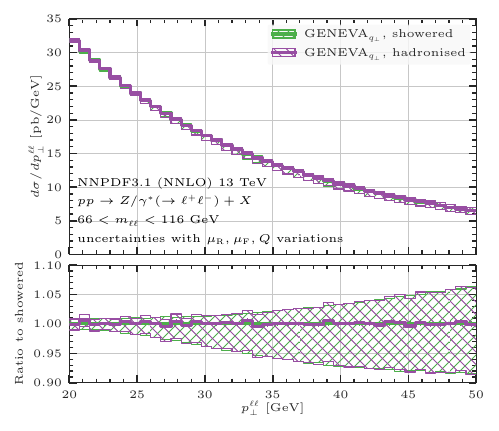}%
  \end{subfigure}
  \hspace*{\hspacebetweenthreeplots}
  \begin{subfigure}[b]{\rescalethreeplots}
    \includegraphics[width=\textwidth]{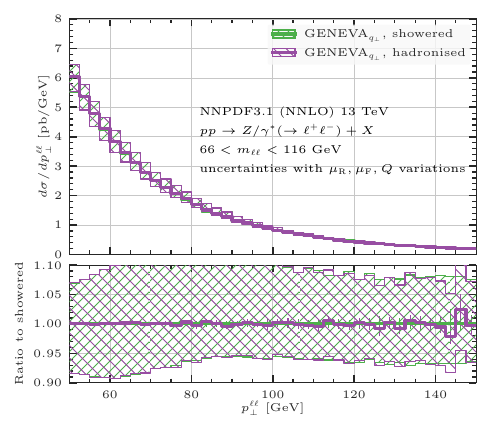}%
  \end{subfigure}
  \vspace{\spacebeforefigurecaption}
  \vspace{\spacebeforefigurecaption}
  \caption{Comparison of the $\pt$ spectra between the partonic N$^3$LL+NLO$_1$ and the showered results, after interfacing to \pythiaEight, before the inclusion of nonperturbative effects (above), and after hadronisation and MPI (below). The peak (left), transition (centre) and tail (right) regions are shown.}
	\label{fig:nnlopscmp2}
\end{figure*}

The showering of 1- and 2-jet events is more delicate.
As discussed in \Sec{geneva}, the separation between 1- and 2-jet events is achieved by means of a Sudakov form factor $U_1(\Phi_1, \mathcal T_1)$, which suppresses the \onejet cross section at small values of $\mathcal T_1$.
The mapping used preserves the \zerojet resolution parameter $\Tau_0$, i.e.~$\mathcal T_0$ in the original \geneva implementation and $\pt$ in this work.
Even with $\mathcal T_1^{\rm cut} = 1$~GeV, $\sigma_{1}^{\mathrm{MC}}$ remains sizeable and must be handled with a certain care.
In the original \geneva implementation, it was necessary to reduce its size to avoid further distortions of the $\mathcal T_0$ distribution after the shower.
Here we prefer to follow the same approach, which also helps to reduce the size of the nonsingular contributions in \Eq{nonsing2jet}.
Properly, to fully account for configurations where \mbox{$\pt \sim \mathcal T_1 \ll
Q$}, one should extend the resummation framework used here and include a joint
resummation for the two jet resolutions. Unfortunately, this is not yet
available at the required logarithimic order.\footnote{Results for the joint
  resummation at lower orders or for other observables are 
  e.g.~available in~\cite{Larkoski:2014tva,Procura:2014cba,Lustermans:2019plv,Monni:2019yyr}}
In the \mbox{$\mathcal T_1 \ll \pt$} limit it is however possible to approximate the
correct behaviour and suppress the size of the \onejet cross section $\sigma_1^{\rm MC}$ by multiplying it by an additional LL Sudakov form factor $U_1(\mathcal T^\cut_1,\Lambda_1)$, see \citeRef{Alioli:2019qzz}.
The new \onejet exclusive cross section becomes
\begin{widetext}
\begin{align}
\frac{\mathrm{d} \sigma_{1}^{\mathrm{MC}}}{\df\Phi_{1}} (\Tau_0 > \Tau_0^\cut; \mathcal T_1^\cut; \Lambda_1) =&\; \frac{\dsigMC_{1}}{\df\Phi_{1}} (\Tau_0 > \Tau_0^\cut; \mathcal T_1^\cut) \, U_1(\mathcal T_1^\cut, \Lambda_1)\, ,
\end{align}
while the 2-jet inclusive reads
\begin{align}
\frac{\mathrm{d} \sigma_{\geq 2}^{\mathrm{MC}}}{\df\Phi_{2}} (\Tau_0 > \Tau_0^\cut; \mathcal T_1^\cut,\mathcal T_{1}>\Lambda_1) =&\; \frac{\dsigMC_{\geq 2}}{\df\Phi_{2}} (\Tau_0 > \Tau_0^\cut, \mathcal T_{1}>\mathcal T_{1}^\cut) \nn \\
&+ \frac{\df}{\df \mathcal T_1} \, \frac{\mathrm{d} \sigma^{\mathrm{MC}}_{1}}{\df\Phi_{1}} (\Tau_0 > \Tau_0^\cut; \mathcal T_1^\cut, \mathcal T_1)
\times\cP(\Phi_2) \,  \theta( \mathcal T_1 > \Lambda_1)
\,.
\end{align}
\end{widetext}
The parameter $\Lambda_1$ is set to be much smaller than $\mathcal T_1^{\rm cut}$ and constitutes the ultimate \onejet resolution cutoff (in the present work, we set this cutoff to 0.0001 GeV).
The \onejet cross section is therefore extremely suppressed and accounts for a negligible fraction of the cross section, at the few permille level.

\begin{figure*}[t]
  \begin{subfigure}[b]{\rescaletwoplots}
    \includegraphics[width=\textwidth]{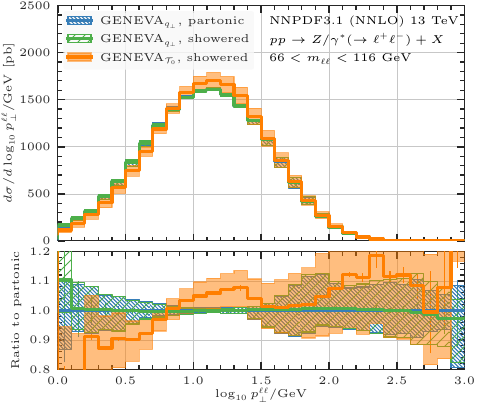}%
  \end{subfigure}
   \hspace*{\hspacebetweentwoplots}
  \begin{subfigure}[b]{\rescaletwoplots}
    \includegraphics[width=\textwidth]{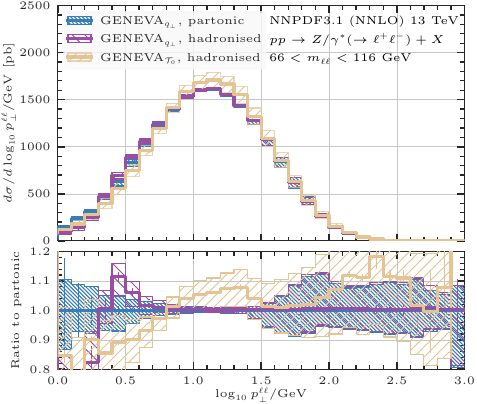}%
  \end{subfigure}
  \vspace{\spacebeforefigurecaption}
  \caption{Comparison between parton level and showered results (left) and between parton level and results after shower, hadronisation and MPI (right) for the transverse momentum distribution of the dilepton pair using \genevapt and \genevatau.}
	\label{fig:res0cmp1}
\end{figure*}

As a result, the inclusive 2-jet cross section accounts for almost all events.
The starting scale of the shower for this set of events is chosen to be sufficiently large to completely fill the phase space available to the $k_\perp$-ordered shower (in particular, in the present work we choose the largest relative $k_\perp$ with respect to the direction of the sister parton for the second emission). The events are then showered, vetoing events which have a $\mathcal T_2$ value larger than the initial $\mathcal T_1$ value once showering is complete.
Events which do not pass this veto are reshowered again until they do.
This in practice corresponds to the implementation of a truncated-vetoed shower~\cite{Nason:2004rx} and thus preserves the LL accuracy of the shower for observables other than $\pt$.

Having discussed the matching conditions of the shower, we now move to discuss the results.
We match the NNLO calculation to the \pythiaEight.245 parton shower and we use the general-purpose CMS MonashStar Tune (\mbox{\texttt{Tune:pp = 18}}).
As previously discussed, it would be desirable to maintain an agreement between the precise parton level predictions and the showered result.
This is possible by choosing a more local recoil scheme in \pythia, through
the option \texttt{SpaceShower:dipoleRecoil = on}, which affects the colour-singlet
kinematics less compared to the standard recoil scheme~\cite{Cabouat:2017rzi,Monni:2020nks}.

This scheme does not change the transverse momentum of the colourless system
when there is an emission involving an initial-final dipole and it was shown to improve the agreement between parton level and showered predictions for the transverse momentum of a diphoton pair in our previous work~\cite{Alioli:2020qrd}.
We stress that the choice of the recoil scheme has important implications on the accuracy of parton showers~\cite{Nagy:2009vg,Dasgupta:2018nvj,Bewick:2019rbu,Dasgupta:2020fwr,Forshaw:2020wrq,Hamilton:2020rcu} and might induce spurious effects at NLL. Nonetheless, since in this work we assume that the accuracy of our predictions is dictated by the parton shower we leave a study of the formal accuracy to future work.

\begin{figure*}[t]
  \begin{subfigure}[b]{\rescaletwoplots}
    \includegraphics[width=\textwidth]{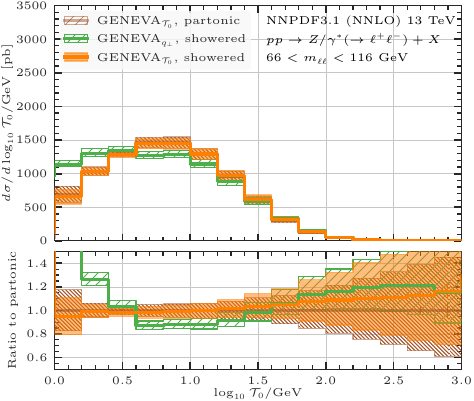}%
  \end{subfigure}
   \hspace*{\hspacebetweentwoplots}
  \begin{subfigure}[b]{\rescaletwoplots}
    \includegraphics[width=\textwidth]{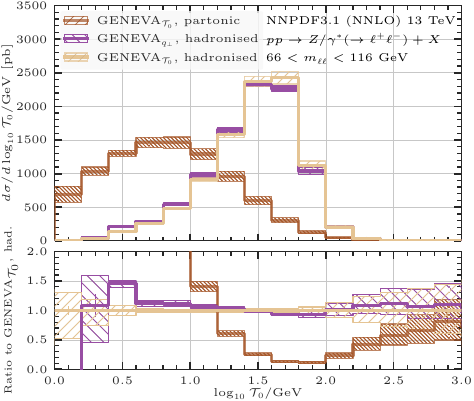}%
  \end{subfigure}
  \vspace{\spacebeforefigurecaption}
  \caption{Same as \Fig{res0cmp1}, now comparing the $\mathcal T_0$ distribution.}
	\label{fig:res0cmp2}
\end{figure*}

\begin{figure*}[t]
  \begin{subfigure}[b]{\rescaletwoplots}
    \includegraphics[width=\textwidth]{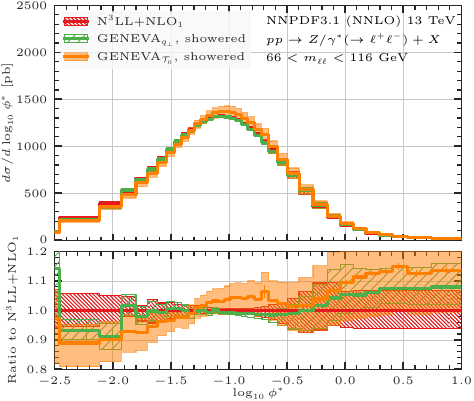}%
  \end{subfigure}
   \hspace*{\hspacebetweentwoplots}
  \begin{subfigure}[b]{\rescaletwoplots}
    \includegraphics[width=\textwidth]{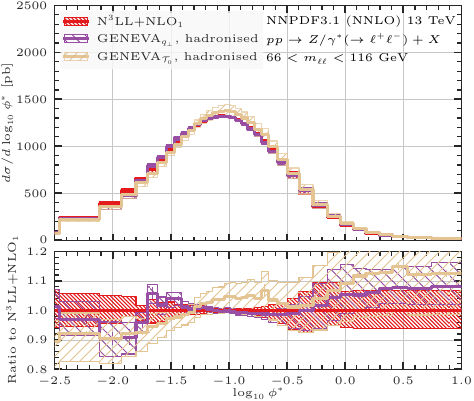}%
  \end{subfigure}
  \vspace{\spacebeforefigurecaption}
  \caption{Comparison between the N$^3$LL+NLO$_1$ prediction and showered results (left) and between the N$^3$LL+NLO$_1$ prediction and results after hadronisation and MPI (right) for the $\phi^*$ distribution using \genevapt and \genevatau.}
	\label{fig:res0cmp3}
\end{figure*}

Nonperturbative effects are another possible source of modification of the $\pt$ distribution induced by the matching with a parton shower.
In general, the values of the nonperturbative parameters in \pythiaEight have been tuned starting from predictions which were lacking
the higher-order effects that are instead included in this work.
Therefore, it is not clear whether their values reflect real nonperturbative effects or simply the lack of perturbative ingredients.
When matching to an NNLO computation it would be advisable to investigate this and eventually perform a retuning.
While the determination of an optimal tune is beyond the scope of this work, we observed that the description of the spectrum in the small $\pt$ region strongly depends on the value of the intrinsic (or primordial) $k_\perp$ transverse momentum of the incoming partons, determined by the \pythia option \texttt{BeamRemnants:primordialKThard}.
In the CMS MonashStar tune the value for this parameter is set to 1.8 GeV; nonetheless, we observed that this value is responsible for a shift in the spectrum outside scale variation bands up to values of $\pt\sim 10$ GeV, i.e.~in a region where one would naively expect nonperturbative effects to play a minor role.
Therefore, we preferred to lower this value to $\sim 0.6$ GeV, such that the uncertainty bands before and after the shower overlap.
In the following, we shall use this value as our default.

The effect of changing the recoil scheme is shown in \Fig{showercmp} for the rapidity distribution of the dilepton pair and for the transverse momentum of the vector boson.
In the plots we compare the parton level predictions with the results after
the shower with different recoil schemes, in the absence of hadronisation and multi particle interactions (MPI) effects.
The difference on the rapidity distribution is negligible almost everywhere:
the dipole recoil scheme is indistinguishable from the parton level
results, while there are tiny discrepancies for the default recoil scheme, but
these appear only at very large rapidities.
On the other hand, the effect on the transverse momentum distribution is somewhat more pronounced, although the two schemes are in good agreement within the scale uncertainty bands, with differences at the few percent level.
In particular, when the more local recoil scheme is chosen, the agreement between the showered result and the N$^3$LL+NLO$_1$ result at the parton level is excellent across the whole spectrum.

We now compare the partonic, showered and hadronised result for a selected set of distributions.
In order to keep the analysis simple, we do not include QED effects in the showered results.
In the following, we always include MPI effects in our hadronised results.
We start by presenting this comparison in \Fig{nnlopscmp1} for the rapidity distribution of the lepton pair and for the transverse momentum of the hardest lepton.
On the left panel, we observe that for the rapidity distribution the NNLO accuracy is mantained
at a very precise level after the showering and the hadronisation stages, as
should be expected by the inclusive nature of this distribution.
On the right panel, the same excellent agreement is also seen in the case of the $p_{T,\ell,\rm
  hardest}$ distribution case, with differences well inside statistical fluctuations.

Next, in~\Fig{nnlopscmp2} we focus on the transverse momentum distribution of the
lepton pair, comparing the partonic result to the result after showering (upper
row) and
after hadronisation and MPI (lower row). We present three separate plots, in the peak, transition
as well as in the tail region  of the distribution.
An extremely good agreement between the showered and the partonic results is
observed in all three regions, both for the central predictions and for
the scale uncertainty bands. When the hadronisation (and MPI) stage is also
added,  some differences arise, localised in the small $\pt$ region. There we observe that  the peak becomes more
pronounced and the spectrum 
is more suppressed at small $\pt$, while at large $\pt$ the showered and
hadronised results are again in agreement, as one expects from factorization.

\begin{figure*}[t]
  \begin{subfigure}[b]{\rescaletwoplots}
    \includegraphics[width=\textwidth]{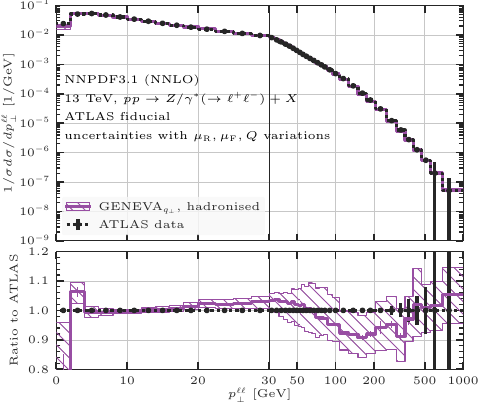}%
  \end{subfigure}
  \hspace*{\hspacebetweentwoplots}
  \begin{subfigure}[b]{\rescaletwoplots}
    \includegraphics[width=\textwidth]{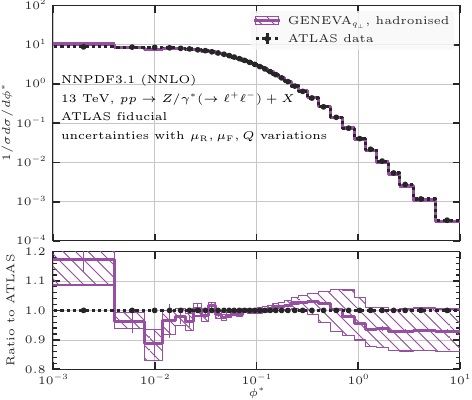}%
  \end{subfigure}
  \vspace{\spacebeforefigurecaption}
  \caption{Comparison between \geneva predictions and the ATLAS data for the transverse momentum distribution (left) and for the $\phi^*$ observable (right).}
	\label{fig:datacomp1}
\end{figure*}

\begin{figure*}[t]
  \begin{subfigure}[b]{\rescaletwoplots}
    \includegraphics[width=\textwidth]{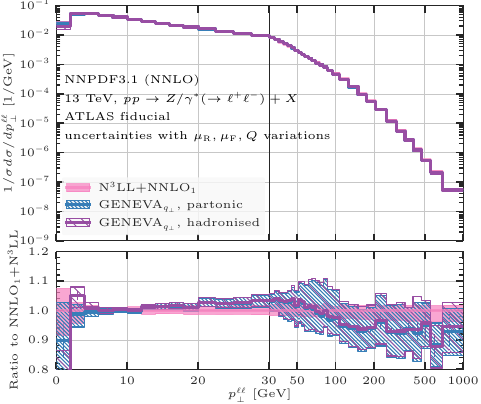}%
  \end{subfigure}
   \hspace*{\hspacebetweentwoplots}
  \begin{subfigure}[b]{\rescaletwoplots}
    \includegraphics[width=\textwidth]{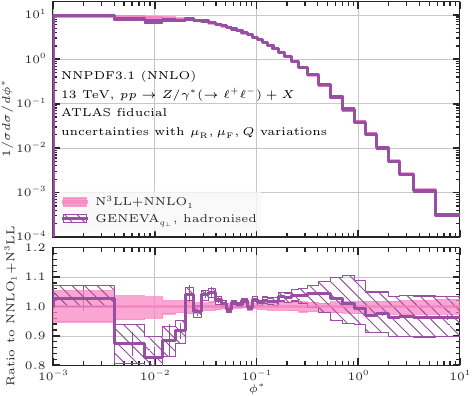}%
  \end{subfigure}
  \vspace{\spacebeforefigurecaption}
  \caption{Comparison between \geneva results and N$^3$LL+NNLO$_1$ predictions for the transverse momentum distribution (left) and for the $\phi^*$ observable (right) within ATLAS fiducial cuts.}
	\label{fig:datathcmp}
\end{figure*}

\subsection{Comparison of resolution variables}

The implementation of two different \zerojet resolution variables opens up the possibility of comparing the predictions after shower, hadronisation and MPI obtained using \geneva in the two cases.
Moreover, we can also compare the results to the parton level predictions, where the higher-order resummation is guaranteed by construction for each \zerojet resolution variable.
We show this comparison in \Fig{res0cmp1} and \Fig{res0cmp2} for the $\pt$ and
$\mathcal T_0$ spectra, respectively. There, the N$^3$LL+NLO$_1$
(NNLL$^\prime$+NLO$_1$) results at the parton level for each \zerojet
resolution variable  are compared to the results after showering and after
hadronisation obtained using $\pt$ and $\mathcal T_0$ in \geneva.
For the transverse momentum distribution we observe differences at the $10-20\%$ level between the results obtained with \genevapt and \genevatau after the shower up to values of $\pt$ close to 30 GeV, though the uncertainty bands always overlap.
The \genevatau result is more suppressed at small $\pt$ and slightly harder above the peak.
The differences are reduced at larger values of $\pt$, although the \genevatau spectrum is somewhat harder than that of \genevapt.
A similar pattern is visible after hadronisation and MPI.

Analogous differences can be observed when comparing the $\mathcal T_0$ spectra obtained using the two different implementations.
The showered results are in good agreement for values of $\mathcal T_0$ larger than 30 GeV, while they start to differ below.
The differences are as large as $50\%$ at small values of $\mathcal T_0$, where the \genevapt result is considerably harder than the NNLL$^\prime$+NLO$_1$ result.
The hadronisation and especially the addition of MPI effects significantly modify the $\mathcal T_0$ distribution, and brings the two results closer also at small values of $\mathcal T_0$.
To accommodate  this larger difference, instead of the usual ratio with respect to the partonic result, in the right panel of \Fig{res0cmp2} we show the ratio with respect to the \genevatau result after hadronisation and MPI. 

Finally, we show the same comparison for the $\phi^*$ variable~\cite{Banfi:2010cf} in \Fig{res0cmp3}.
Although this observable is not fully resummed at N$^3$LL accuracy in the
\genevapt implementation, we expect to observe good agreement between the \genevapt and the resummed results since the two observables are closely related.
Indeed we observe that after showering the \genevapt result is close to the
N$^3$LL+NLO$_1$ result obtained with \Matrix{}+\radish. The \genevatau result
displays instead differences similar to those seen for the $\pt$ distribution, both before and after hadronisation.

\begin{figure*}[t]
  \begin{subfigure}[b]{\rescaletwoplots}
    \includegraphics[width=\textwidth]{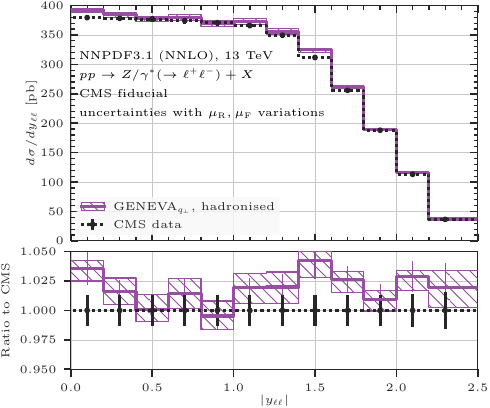}%
  \end{subfigure}
  \hspace*{\hspacebetweentwoplots}
  \begin{subfigure}[b]{\rescaletwoplots}
    \includegraphics[width=\textwidth]{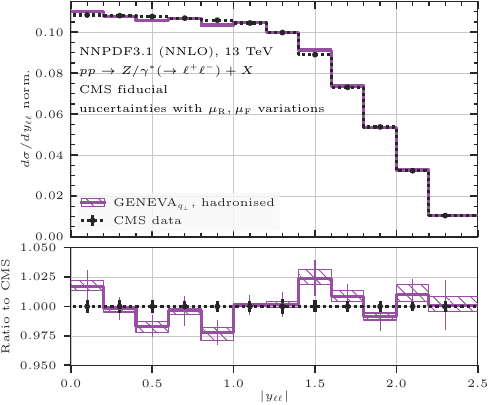}%
  \end{subfigure}
  \begin{subfigure}[b]{\rescaletwoplots}
    \includegraphics[width=\textwidth]{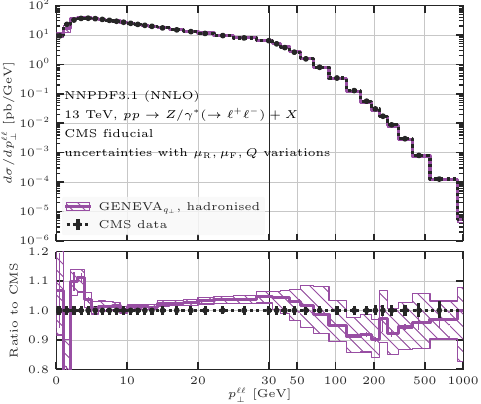}%
  \end{subfigure}
  \hspace*{\hspacebetweentwoplots}
  \begin{subfigure}[b]{\rescaletwoplots}
    \includegraphics[width=\textwidth]{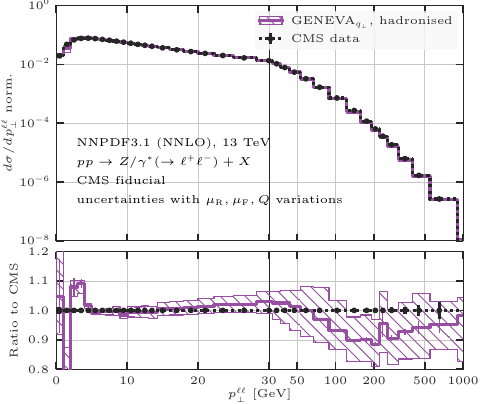}%
  \end{subfigure}
  \begin{subfigure}[b]{\rescaletwoplots}
    \includegraphics[width=\textwidth]{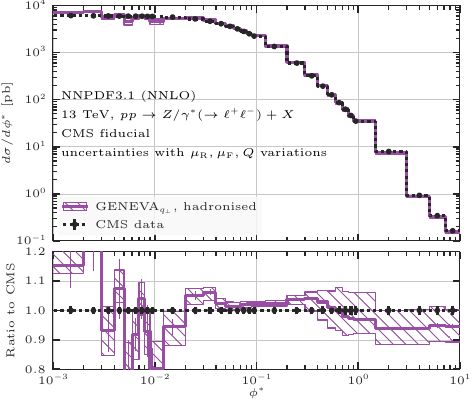}%
  \end{subfigure}
  \hspace*{\hspacebetweentwoplots}
  \begin{subfigure}[b]{\rescaletwoplots}
    \includegraphics[width=\textwidth]{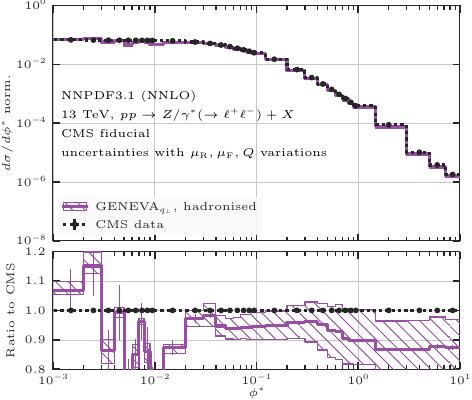}%
  \end{subfigure}
  \vspace{\spacebeforefigurecaption}
  \caption{Comparison between \geneva and the CMS data for different observables. Normalised distributions are shown on the right, see text for details.}
	\label{fig:datacomp2}
\end{figure*}


\begin{figure*}[t]
  \begin{subfigure}[b]{\rescaletwoplots}
    \includegraphics[width=\textwidth]{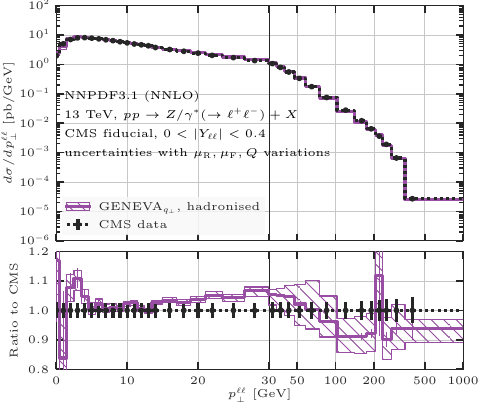}%
  \end{subfigure}
  \hspace*{\hspacebetweentwoplots}
  \begin{subfigure}[b]{\rescaletwoplots}
    \includegraphics[width=\textwidth]{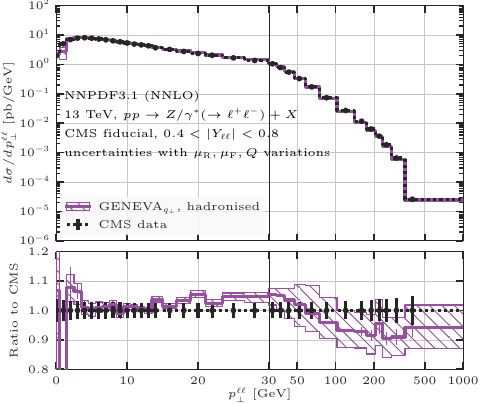}%
  \end{subfigure}
  \begin{subfigure}[b]{\rescaletwoplots}
    \includegraphics[width=\textwidth]{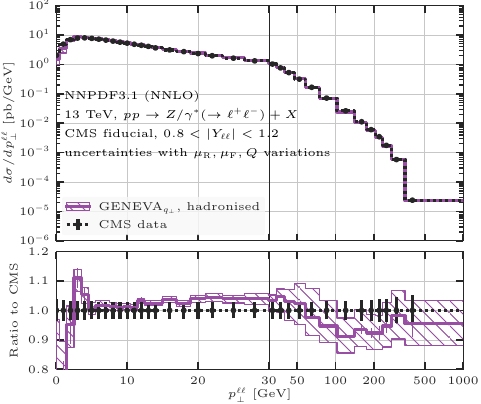}%
  \end{subfigure}
  \hspace*{\hspacebetweentwoplots}
  \begin{subfigure}[b]{\rescaletwoplots}
    \includegraphics[width=\textwidth]{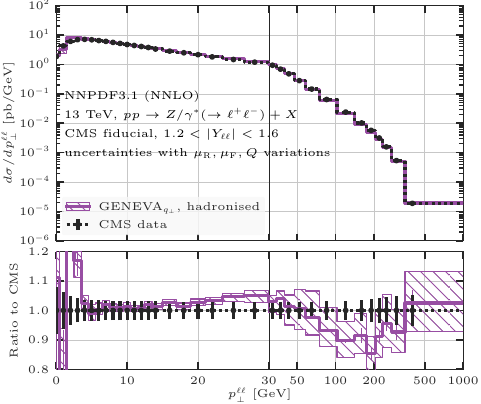}%
  \end{subfigure}
  \vspace{\spacebeforefigurecaption}
  \caption{Comparison between \geneva and the CMS data for the $\pt$ distribution in different rapidity slices.}
	\label{fig:datacomp3}
\end{figure*}

\section{Results and comparison to LHC data}\label{sec:result}

In this section we compare our predictions against 13~TeV data collected at the LHC by the ATLAS~\cite{Aad:2019wmn} and the CMS~\cite{Sirunyan:2019bzr} experiments.
We have generated events using the same settings as detailed in \Sec{val1} regarding the choice of PDF and central scales, and we use the shower settings as described in \Sec{shower}.
We remind the reader that our predictions thus include hadron decay and MPI effects, but do not include QED shower effects.

\subsection{Comparison to ATLAS data}

We start by showing the comparison of our predictions with the ATLAS data of \citeRef{Aad:2019wmn}.
The $Z$ boson is reconstructed by selecting the two hardest same-flavour
opposite-sign (SFOS) leptons in the final state.
We then apply the following cuts to our events:
\begin{equation}
	p_\perp^\ell > 27\, {\rm GeV}, \quad |\eta_\ell| < 2.47, \quad m_{\ell \ell} \in [66,116]\, \rm{GeV}.
\end{equation}
In \Fig{datacomp1} we compare our predictions for the normalised $\pt$
and $\phi^*$ distributions.
For $\pt$, we use a scale which is linear up to $30$~GeV and logarithmic for larger values.
The former are in very good agreement with the data in the whole $\pt$ range.
Below $30$~GeV, the central prediction is within a few percent of the data, and only in the first two bins, where hadronisation and nonperturbative effects play a prominent role, do the scale uncertainty bands fail to cover the experimental data.
Our predictions are also in good agreement with the $\phi^*$ measurements,
matching them within scale uncertainty bands  down to values of $\phi^*\sim
0.01$; at lower values the differences reach the $20\%$ level in the first bin,
and the perturbative uncertainty does not cover the data. Here, the inclusion
of shower and nonperturbative uncertainties as well as the development of a
dedicated tuning could help ameliorate the agreement.

Finally, in \Fig{datathcmp} we can compare our predictions with parton level results at N$^3$LL+NNLO$_1$ accuracy~\cite{Bizon:2019zgf} for the $\pt$ and $\phi^*$ distributions obtained by matching \radish results with fixed-order predictions from NNLOJET~\cite{GehrmannDeRidder:2005cm,Daleo:2006xa,Currie:2013vh} in the ATLAS fiducial region.
In the $\pt$ case we also show the \geneva parton level predictions at N$^3$LL+NLO$_1$ accuracy for reference.
Our results for the \pt spectrum are in good agreement with the parton level predictions across the whole range.
In particular, the results are within a few percent from the N$^3$LL+NNLO$_1$ result up to $30$ GeV; at larger values, where NNLO$_1$ corrections become dominant, the differences can reach $\mathcal O(10\%)$.
We observe a similar agreement in the $\phi^*$ case, with differences at most at the $10\%$ level around $\phi^*\sim 0.01$.

\subsection{Comparison to CMS data}

Finally, we compare our predictions with the CMS data of \citeRef{Sirunyan:2019bzr}.
The fiducial region in this case is defined by the selection cuts
\begin{align}
	p_\perp^\ell > 25\, {\rm GeV}, \qquad |\eta_\ell| < 2.4, \\ |m_{\ell \ell}-91.1876\, \rm{GeV}| < 15\, \rm{GeV}.
\end{align}
We compare our predictions both at the absolute level and for normalised distributions.
We note that in the analysis of \citeRef{Sirunyan:2019bzr} the
latter are defined by dividing by the sum of the weights rather than
by the integrated cross section, i.e. it is not normalised by the fiducial cross section.

In \Fig{datacomp2} we show the comparison for the absolute rapidity
$|y_{\ell \ell}|$ and for the $\pt$ and $\phi^*$
distributions.
In the first row we observe a reasonably good agreement for the rapidity distribution.
The theoretical predictions are a few percent higher than the experimental data at the absolute level, and oscillate around the data in the normalised distribution.
In the central row, the predictions for $\pt$ are also in good agreement with the CMS
data, repeating the same pattern already observed when comparing to ATLAS
data.  This is especially true at the normalised level.

In the last row, we observe that the theoretical predictions display statistical fluctuations
more pronounced than in the ATLAS case. This is due to the very fine binning at small values of
$\phi^*$.
Moreover, the normalised predictions seem to consistently undershoot the data.
A possible explanation for this effect could lie in the interplay between the statistical
fluctuations at small values of $\phi^*$ and the atypical normalisation chosen
for this particular analysis, which enhances the impact of the low $\phi^*$
region across the whole spectrum.
This can be further appreciated by noticing that the relative size of the
uncertainty bands is somewhat larger in the normalised $\phi^*$ distribution,
defying a na\"ive expectation.
Finally, in \Fig{datacomp3} we compare our predictions for the transverse
momentum distribution in different rapidity slices with the CMS data.
In all cases, we observe a good agreement, with larger differences localised in the first few bins where hadronisation effects are more significant.

\section{Conclusion}\label{sec:conclusion}

In this work we have presented an extension of the \geneva method which uses
the transverse momentum of the colour singlet as the \zerojet resolution
variable. 
As a first application, we have provided an NNLO+PS description of Drell-Yan pair production, by matching the parton level calculation with the \pythiaEight parton shower.
The method presented here is fully general and can be readily applied to other colour-singlet processes.

In this specific implementation, the transverse momentum resummation is
performed at N$^3$LL accuracy by interfacing \geneva with the \radish code.
We have validated our predictions of the transverse momentum resummation by comparing  with the N$^3$LL+NNLO$_0$ results obtained with the \Matrix{}+\radish interface.
The use of the transverse momentum as \zerojet resolution variable allowed us
to reduce the impact of missing power corrections in the NNLO calculation.
Setting $\pt^\text{cut} = 1$~GeV, we have found that the  missing power corrections
contribute below the permille level. We validated our NNLO differential
predictions by comparing a few selected distributions against the \Matrix program.
The availability of a fully exclusive event generator at this accuracy allows for the evaluation of any fiducial cross section, also correctly resumming linear power corrections associated with cuts on the lepton kinematics.

We have then studied the impact of parton shower and hadronisation on our
predictions and found that the NNLO description of inclusive observables is
preserved after both shower and hadronisation.
An important outcome of this study is the observation that it is possible to
maintain an extremely good agreement between the N$^3$LL+NLO$_1$ result for
the transverse momentum spectrum and the \geneva results after the shower by
choosing a more local  recoil scheme in \pythiaEight.
We have also quantified the impact of nonperturbative corrections after hadronisation on the transverse momentum distribution, finding, as expected, that they are localised below the peak.

The construction of an NNLO+PS event generator is subject to several assumptions, and the choice of which resolution variables are used is of particular importance.
The availability of two different \zerojet resolution variables within the \geneva framework allows us to robustly assess the impact of such choices on differential observables for the first time.
We have studied the effect of the shower and hadronisation (including MPI effects) on the higher-logarithmic partonic predictions using both $\mathcal T_0$ and $\pt$ as \zerojet resolution variable.
The major
differences are observed below the peak region, where the higher-order
resummation of the correct logarithms is necessary to reproduce the correct
behaviour. Nonetheless, the two predictions are in  reasonable agreement,
with the size of the differences never exceeding $15\%$ for the $\pt$ and $\phi^*$ distributions. 

Finally, we compared our predictions, including hadronisation and MPI effects,
to LHC data at 13 TeV. A good agreement has been observed both for inclusive
and for more exclusive distributions, such as  the transverse momentum spectrum of the dilepton system as well as the $\phi^*$ variable, across the whole spectrum.
The description at very small $\pt$ and $\phi^*$ is sensitive to the details of the hadronisation procedure, suggesting that dedicated studies are needed in order to determine the optimal value of the nonperturbative parameters in a NNLO+PS computation.
It should be emphasised, however, that at this level of precision, the
inclusion of EW corrections and a careful study of the impact of mass effects
become relevant. We have refrained from including these effects in this study,
leaving them to future work.
Another direction in which our predictions could be improved concerns the
inclusion of higher-order matrix elements, beyond what is available in a NNLO
calculation for Drell-Yan with no additional jets.

This work could be expanded even further by extending the study to the use of
alternative \onejet resolution variables and their resummation, as well as the
interplay between the \zerojet and \onejet resummations.
In particular, in view of the recent advancements in the development of parton
showers beyond LL accuracy, the choice of a suitable \onejet resolution variable might prove crucial in order to ensure that the matching of NNLO calculations to parton shower preserves the shower accuracy.

The code used to produce the results presented in this work is available upon request from the authors, and will be made public in a future \geneva release.

\noindent {\bf Acknowledgements.}

We are grateful to P.~Monni, P.~Nason and F.~Tackmann for constructive comments on the manuscript.  
SA is also grateful to P.~Nason for clarifying discussions about the parton shower interface.
LR thanks P.~Monni, P.~Torrielli and E.~Re for discussions.
We thank A.~Apyan for clarifications concerning the CMS analysis of \citeRef{Sirunyan:2019bzr}.
The work of SA, AB, AG, SK,
RN, DN and LR is supported by the ERC Starting Grant REINVENT-714788.
SA and ML acknowledge funding from Fondazione Cariplo and Regione
Lombardia, grant 2017-2070.  The work of SA is also supported by MIUR
through the FARE grant R18ZRBEAFC.
The work of LR is also supported by the Swiss National Science Foundation (SNF) under contract 200020\_188464.
CWB is supported by the Director, Office of Science, Office of
High Energy Physics of the U.S. Department of Energy under the
Contract No. DE-AC02-05CH11231.
We acknowledge the CINECA award
under the ISCRA initiative and the National Energy Research Scientific
Computing Center (NERSC), a U.S. Department of Energy Office of
Science User Facility operated under Contract No. DEAC02-05CH11231,
for the availability of the high performance computing resources
needed for this work.
LR would like to thank LBNL and UC Berkeley for the kind hospitality and for financial support during stages of this work.

\appendix

\section{$2\rightarrow 1$ mapping for final-state radiation}\label{sec:mapping}

In this appendix we describe the $\vec q_T$-preserving mapping used for the implementation of the $\mathcal T_1$
resummation.
We define the phase space for the emission of $N$ particles by also including the momentum fractions $x_a$, $x_b$ of the two incoming partons, such that
\begin{align}
	d \Phi_N (x_a, & x_b; p_1, \ldots p_N) \equiv \nonumber \\
	&  dx_a dx_b d {\boldsymbol \Phi}_N (x_a P_a + x_b P_b; p_1, \ldots p_N)
	\,,
\end{align}
where
\begin{align}
	d {\boldsymbol \Phi}_N (Q; p_1, \ldots p_N) =& (2 \pi)^4 \delta(Q - \sum p_i) \nonumber
	\\ &\times \prod_{i=1}^N \frac{d^4 p_i}{(2 \pi)^3} \delta(p_i^2 -m_i^2) \theta(p_i^0)
\end{align}
is the $N$-body Lorentz-invariant phase space. The momenta of the incoming
hadrons are
\begin{equation}
	P_a = E_{\rm cm} \frac{n_a}{2},\quad P_b = E_{\rm cm} \frac{n_b}{2}\,,
\end{equation} where $E_{\rm cm}$ is the hadronic centre-of-mass energy.
For later convenience, we have introduced  light-cone coordinates relative to the beam axis as
\begin{equation}
	n^\mu \equiv n_a^\mu = (1, 0, 0, 1) \,, \quad \bar n^\mu \equiv n_b^\mu = (1, 0, 0, -1)
	\,,
\end{equation}
such that
\begin{align}
	P_a^- =  P_b^+ & {} =  E_{\rm cm}, \qquad P_a^+ = P_b^- = 0, \\ \nonumber
	& (P_a + P_b)^2 = E_{\rm cm}^2
	\,.
\end{align}
The momenta of the incoming partons are then
\begin{align}
p_a &= x_a P_a,\quad p_a^- = x_a E_{\rm cm},\quad p_a^+ = 0
\nonumber\\
p_b & = x_b P_b,\quad p_b^+ = x_b E_{\rm cm},\quad p_b^- = 0\,,
\end{align}
and we define
\begin{equation}
	s = (p_a + p_b)^2 = x_a x_b E_{\rm cm}^2
	\,.
\end{equation}

The phase space projection is a variable transformation from the $N+1$ phase space onto the underlying Born phase space $\bar \Phi_N$, such as
\begin{align}
	&\Phi_{N+1} (x_a, x_b; p_1,  \ldots p_N, p_{N+1})
	 \leftrightarrow \nonumber \\
	& \qquad \{ \bar \Phi_{N} (\bar x_a, \bar x_b;  \bar p_1, \ldots \bar p_N), \Phi_{\rm rad} (\xi, y, \phi)\}
	\,,
\end{align}
where $\Phi_{\rm rad} (\xi, y, \phi)$ is the phase space of the radiation and $\xi, y, \phi$ are the radiation variables which parametrise the emission.

In particular, we are interested in the case in which $p_1, \ldots , p_M$, with
$M < N$, are the momenta of colourless particle, and
\begin{equation}
	q \equiv \sum_1^M  p_i
\end{equation}
is the total momentum of the colour-singlet system. For the Drell-Yan
production process studied in this paper, we  specifically consider the mapping
\begin{equation}
	\Phi_2 (x_a, x_b; q, p_1, p_2) \leftrightarrow \{  \Phi_1 (\bar x_a, \bar x_b; \bar q, \bar p) , \Phi_{\rm rad} (\xi, y, \phi) \}
	\,,
\end{equation}
where the number of particles in the phase space label has now been reduced to
match the notation adopted in the main text by omitting the
number of leptons from the count.

In order to construct the projection, we start from the relations of momentum conservation
\begin{align}\label{eq:momcons}
 & x_a \Ecm = p_1^- + p_2^-  + q^-,    & \bar x_a\Ecm = \bar p^- + \bar q^-, \nonumber \\
 &   x_b \Ecm = p_1^+ + p_2^+  + q^+,    & \bar x_b \Ecm = \bar p^+ + \bar q^+, \nonumber \\
 & 0 = {\vec{ p}_{1,\perp}} + {\vec{ p}_{2,\perp}} + {\vec{ q}_{\perp}}, & 0 = \vec {\bar p}_{\perp}+\vec {\bar q}_{\perp}
 \,.
\end{align}
Imposing also the conservation of the four-momentum of the vector boson $q$ we have
\begin{align}
	q^- = \bar q^-,  \qquad  q^+=\bar q^+,  \qquad  \vec{ q}_\perp =
        \vec{\bar q}_\perp
	\,.
\end{align}
This immediately gives the relation
\begin{equation}
	\vec{\bar p}_\perp = \vec{p}_{1,\perp} + \vec{p}_{2,\perp}
	\,.
\end{equation}

The projection is uniquely defined once we decide how to map the (massive) jet with momentum
\begin{align}
p_{12} = p_1 + p_2
\end{align}
 into the massless parton $\bar p$ .
We choose to fix  the longitudinal component
\begin{equation}
	\bar p^z = p_1^z + p_2^z
	\,.
\end{equation}
From the on-shell relation
\begin{align}\label{eq:onshellpbar}
	0 = \bar p^2 = (\bar p^0)^2 -(\bar p^z)^2 - (\vec{\bar p}_\perp)^2
\end{align}
one can now easily determine $\bar p^0$.
Having fixed $\bar p^z$ and $\bar p^0$, the barred system and therefore the
complete projection is fully determined, since we can obtain
\begin{align}
	\bar p^{+} &= \bar p^0 - \bar p^z, \qquad \bar p^{-} = \bar p^0 + \bar p^z\,,
\end{align}
and solve \Eq{momcons} for $\bar x_a$ and $\bar x_b$.

To construct the inverse mapping  it is convenient to first reconstruct the total momentum $p_{12}$.
Once $p_{12}$ is known, $p_1$ and $p_2$ can be constructed by decaying $p_{12}$ in its restframe with decay angles $\{ \cos \vartheta, \phi \}$.
Here $\phi$ is the azimuthal radiation angle whilst $\cos \vartheta$ can be determined
by the variable $\xi$ from the relation
\begin{equation}\label{eq:theta}
2 p_2^0 = \xi \Ecm = p_{12}^0 + |\vec p_{12}| \cos \vartheta
\,.
\end{equation}
Note that the $\Phi_{\rm rad}$ variables are defined in the hadronic centre-of-mass frame, i.e.
\begin{align}
	\xi &= \frac{2 p_2^0}{\Ecm} = \frac{2 p_{12}^0 (1-z)}{\Ecm}
	\,, \\
	y &= \frac{\vec p_1 \cdot \vec p_2}{|\vec p_1| |\vec p_2|} = 1 -\frac{p_{12}^2}{2 p_1^0 p_2^0}
	\,.
\end{align}

To reconstruct $p_{12}$ we then must solve the system
\begin{align}
  \begin{cases}
	p_{12}^2 = 2 p_1^0 p_2^0 (1-y) = (2 p_{12}^0 - \xi \Ecm) \, \xi \Ecm \frac{1-y}{2}\\~\\
	 p_{12}^2 = (p_{12}^0)^2 - (p^z_{12}) - q_\perp^2 =  (p_{12}^0)^2 -
    (\bar p^z) - q_\perp^2
    \end{cases}
\end{align}
for $p_{12}^2$ and $p_{12}^0$. One obtains two solutions, but only one of them is physical such that $p_{12}^2 >0$ and $p_{12}^0 >0$:
\begin{align}\label{eq:p12sq}
	p_{12}^2 = & \frac12 \left(-\Ecm^2 \xi^2 y + \Ecm^2 \xi^2 y^2  \phantom{\sqrt{4 (\bar p^z)^2}}  \right. \\
 &  + \, \Ecm \xi \sqrt{4 (\bar p^z)^2 + 4 q_\perp^2 - \Ecm^2 \xi^2 + \Ecm^2 \xi^2 y^2}\nonumber\\
  & \left. - \,  \Ecm \xi y \sqrt{4 (\bar p^z)^2 + 4 q_\perp^2 - \Ecm^2 \xi^2 + \Ecm^2 \xi^2 y^2}\right) ,\nonumber\\
  p_{12}^{0} =&  \frac12 \left(\Ecm \xi (1-y) \phantom{\sqrt{4 (\bar p^z)^2}} \right. \\
  & +\left. \sqrt{
    4 (\bar p^z)^2 + 4 q_\perp^2 - \Ecm^2 \xi^2 + \Ecm^2 \xi^2 y^2}\right) \nonumber
    \,.
\end{align}
This is enough information to fully reconstruct $p_{12}$, since
\begin{align}
	\bar p_{12}^0 = \frac{p_{12}^- + p_{12}^+}{2}
	\,, \qquad \bar p^z = p_{12}^z =\frac{p_{12}^- - p_{12}^+}{2}
	\,,
\end{align}
which allows us to obtain $p_{12}^\pm $.

The Jacobian can be derived starting by the relation between $d \Phi_2 $ and $d \Phi_1$
\begin{align}
	\frac{\df \Phi_2}{\df \Phi_1 \df \Phi_{\rm rad}} =  \frac{1}{\xi \Ecm^2} \frac{\bar p}{p_{12}^0} \frac{\bar q^+}{q^+} \frac{ \df p_{12}^z \, \df^2 \!q_\perp \, \df q^+ \df p_{12}^2 \, \df \!\cos \vartheta \, \df \phi}{\df\bar p^z \, \df^2\bar q_\perp^2 \df \bar q^+ \df \xi \, \df y \, \df \phi}
\,.
\end{align}
Since the mapping is defined such as $q = \bar q$, $p_{12}^z =
\bar p^z$, one has
\begin{align}
	\frac{\df \Phi_2}{\df \Phi_1 \df \Phi_{\rm rad}} =   \frac{1}{\xi \Ecm^2}  \frac{\bar p}{p_{12}^0}  \frac{\df p_{12}^2 \, \df \!\cos \vartheta}{\df \xi \, \df y }
	\,,
\end{align}
which can be calculated by using \Eq{theta} and \Eq{p12sq}.
Since the expression is not particularly compact, we refrain from reporting it here.

Finally, we show how one can write the inverse map in terms of $\mathcal T_1$ and the energy fraction $z$.
In order to do that, one should first express $p_{12}$ as a function of $\mathcal T_1$.
Since
\begin{align}
	\mathcal T_1 = \hat E_{12} - |\hat p_{12}|
	\,,
\end{align}
where $\hat{}$ denotes that the variable is evaluated in the centre-of-mass
frame of the leptonic system, we have
\begin{align}
	\mathcal T_1 = \frac{e^{-Y} p_{12}^- + e^Y p_{12}^+}{2} - \frac12 \sqrt{(e^{-Y} p_{12}^- - e^Y p_{12}^+) + 4 q_\perp^2}
	\,,
\end{align}
where
\begin{equation}
	Y = \ln \sqrt{q^-/q^+}
	\,.
\end{equation}
Since the mapping uniquely fixes $(p_{12}^--p_{12}^+)/2 = p_{12}^z = \bar p^z$, and $p_{12}^2 = p_{12}^+ p_{12}^- - q_\perp^2$, one can fully reconstruct $p_{12}$.
One finds that
\begin{widetext}
\begin{align}
	p_{12}^0 =& \mathcal T_1 \cosh Y + \sqrt{(\bar p^z)^2 + q_\perp^2 -\mathcal T_1^2  +
  \mathcal T_1^2 \cosh^2 Y - 2 \mathcal T_1 \bar  p^z \sinh Y}, \\
  p_{12}^2 =& \mathcal T_1 \left(\mathcal T_1 \cosh 2Y - 2 \bar p^z  \sinh Y
  \nonumber \right. \\ 
  & \left. + \cosh Y \sqrt{4((\bar p^z)^2 + q_\perp^2) + 2 \mathcal T_1 (\mathcal T_1 \cosh 2Y - 4 \bar p_z \sinh Y) -2 \mathcal T_1^2 }\right)
  \,,
\end{align}
\end{widetext}
which allows one to determine $\xi$ and $y$ and to compute the Jacobian $\left( \df
\xi \, \df  y \right) / \left( \df \mathcal T_1 \, \df z \right)$ using
\begin{align}
	& \xi = \frac{2 p_{12}^0 (1-z)}{\Ecm}
	\,,
	\nonumber \\
	& y = 1 -\frac{p_{12}^2}{\xi \Ecm (p_{12}^0-\xi \Ecm/2)}
	\,.
\end{align}
Also in this case, we refrain from reporting the rather lengthy expression for the
Jacobian. The interested reader can  find them well documented  in the \geneva code.

\bibliographystyle{apsrev4-1}
\bibliography{genevapt}
\end{document}